\documentclass[a4paper,11pt]{article}

\usepackage{jheppub} 

\usepackage[T1]{fontenc} 
\usepackage{amsmath}
\usepackage{amsthm}
\usepackage{physics}
\usepackage[compat=1.1.0]{tikz-feynman}
\usepackage[T1]{fontenc} 
\usepackage{tensor}
\usepackage{graphicx}
\usepackage[export]{adjustbox}
\usepackage{slashed}
\usepackage{stackrel,amssymb}
\usepackage{tikz}
\usepackage{mathtools}
\usetikzlibrary{arrows.meta}
\usepackage{comment}
\usepackage{ytableau}
\usepackage{pgfplots}
\usepackage{tikz-3dplot}
\usepackage{tikz}
\usepackage{subcaption}
\usetikzlibrary{calc,fadings,decorations.pathreplacing,shadings}
\usepackage{verbatim}
\usepackage{natbib}
\tdplotsetmaincoords{60}{115}
\pgfplotsset{compat=newest}
\tikzset{>={Latex[scale=1.1]}}
\usetikzlibrary{arrows.meta,
	bending,
	decorations.markings, decorations.text,quotes,angles}
\newcommand{\1}{~}
\newcommand{\ee}{\mathrm{e}}

\newcommand{\vvev}[1]{{\left\langle #1 \right\rangle}}
\makeatletter
\newcommand*{\letterdef@}{}
\newcommand*{\letterdef}[3]{%
	\def\letterdef@##1{\expandafter\newcommand\csname #1\endcsname{#2{##1}}}%
	\@tfor\@tempa :=#3\do{\expandafter\letterdef@\expandafter{\@tempa}}}
\makeatother
\letterdef{c#1} {\mathcal}{ABCDEFGHIJKLMNOPQRSTUVWXYZ} 
\letterdef{rm#1}{\mathrm} {dDeimM} 

\newcommand\pgfmathsinandcos[3]{%
	\pgfmathsetmacro#1{sin(#3)}%
	\pgfmathsetmacro#2{cos(#3)}%
}
\newcommand\LongitudePlane[3][current plane]{%
	\pgfmathsinandcos\sinEl\cosEl{#2} 
	\pgfmathsinandcos\sint\cost{#3} 
	\tikzset{#1/.style={cm={\cost,\sint*\sinEl,0,\cosEl,(0,0)}}}
}

\newcommand\LatitudePlane[3][current plane]{%
	\pgfmathsinandcos\sinEl\cosEl{#2} 
	\pgfmathsinandcos\sint\cost{#3} 
	\pgfmathsetmacro\yshift{\RadiusSphere*\cosEl*\sint}
	\tikzset{#1/.style={cm={\cost,0,0,\cost*\sinEl,(0,\yshift)}}} %
}
\newcommand\NewLatitudePlane[4][current plane]{%
	\pgfmathsinandcos\sinEl\cosEl{#3} 
	\pgfmathsinandcos\sint\cost{#4} 
	\pgfmathsetmacro\yshift{#2*\cosEl*\sint}
	\tikzset{#1/.style={cm={\cost,0,0,\cost*\sinEl,(0,\yshift)}}} %
}

\newcommand\DrawLongitudeArc[4][black]{
	\LongitudePlane{\angEl}{#2}
	\tikzset{current plane/.prefix style={scale=1}}
	\pgfmathsetmacro\angVis{atan(sin(#2)*cos(\angEl)/sin(\angEl))} %
	\pgfmathsetmacro\angA{mod(max(\angVis,#3),360)} %
	\pgfmathsetmacro\angB{mod(min(\angVis+180,#4),360} %
	\draw[current plane,#1,opacity=0.4] (#3:\RadiusSphere) arc (#3:#4:\RadiusSphere);
	\draw[current plane,#1]  (\angA:\RadiusSphere) arc (\angA:\angB:\RadiusSphere);
}%
\newcommand\DrawLatitudeCircle[2][1]{
	\LatitudePlane{\angEl}{#2}
	\tikzset{current plane/.prefix style={scale=#1}}
	\pgfmathsetmacro\sinVis{sin(#2)/cos(#2)*sin(\angEl)/cos(\angEl)}
	\pgfmathsetmacro\angVis{asin(min(1,max(\sinVis,-1)))}
	\draw[current plane] (\angVis:1) arc (\angVis:-\angVis-180:1);
	\draw[current plane,opacity=0.4] (180-\angVis:1) arc (180-\angVis:\angVis:1);
}

\tikzset{%
	>=latex, 
	inner sep=0pt,%
	outer sep=2pt,%
	mark coordinate/.style={inner sep=0pt,outer sep=0pt,minimum size=3pt,
		fill=black,circle}%
}

\title{Into the wedge of $\mathcal{N}=2$ superconformal gauge theories}

\author[a]{L. Griguolo,}
\author[b]{L. Guerrini,}
\author[a]{A. Testa,}

\affiliation[a]{Dipartimento SMFI, Universit\`a di Parma and INFN Gruppo Collegato di Parma, \\ Viale G.P. Usberti 7/A, 43100 Parma, Italy}
\affiliation[b]{Faculty of Physics, University of Warsaw, ul. Pasteura 5, 02-093 Warsaw, Poland} 

\emailAdd{luca.griguolo@unipr.it, Luigi.Guerrini@fuw.edu.pl, alessandro.testa@unipr.it}

\abstract{We study $\frac{1}{4}$-BPS Wilson loops in four-dimensional SU$(N$) ${\cal N}=2$ super-Yang-Mills theories with conformal matter in an arbitrary representation $\cR$. These operators are formed of two meridians on the two-sphere separated by an arbitrary opening angle. We conjecture that these observables are encoded in a modification of Pestun's matrix model. 	The matrix representation of these operators 
	resembles
	 that of the $\frac12$-BPS circular Wilson loop, differing only for  a rescaling in the exponent. 
	 We compare the matrix model predictions  with an explicit three-loop calculation in flat space based on standard Feynman-diagram techniques, finding perfect agreement.
Finally, exploiting the matrix model representation of these Wilson loops, we study the large-$N$ limit at strong coupling of  $\mathcal{N}=2$ superconformal QCD, finding a surprising transition in the vacuum expectation value for a critical opening angle.}

\begin{document}
	\maketitle
	\flushbottom
	\section{Introduction and summary of the results}
	The study of supersymmetric Wilson loops in four-dimensional gauge theories with extended supersymmetry algebra plays a crucial role, especially in the AdS/CFT correspondence \cite{Maldacena:1997re, Maldacena:1998im, Rey:1998ik, Aharony:1999ti}, where these operators are dual to certain string configurations and provide an effective probe to study holographic principles. For sufficiently protected operators,   it is possible to compute them  for arbitrary value of the coupling constant by different techniques, such as supersymmetric localization \cite{Pestun:2016zxk}. In $\mathcal{N}=4 $ super-Yang-Mills theories, the paradigmatic examples are the operators constructed by Zarembo \cite{Zarembo:2002an} and the $\frac 12$-BPS circular Wilson loops. The latter are described by a non-trivial function of the gauge coupling which arises from the resummation of ladder-like (non-interacting) Feynman diagrams via a Gaussian matrix model \cite{Pestun:2007rz, Erickson:2000af,Drukker:1999zq,Drukker:2000rr}.
	
	 Reducing the number of supercharges which commute with the operators, the corresponding observables can receive  quantum corrections and their calculation becomes more involved. In $\mathcal{N}=4$ SYM, this scenario was considered in \cite{Drukker:2007qr}, where the authors constructed families of BPS Wilson loops that preserve fewer supercharges than the circular configurations. In these examples supersymmetric localization is still applicable \cite{Pestun:2009nn,Bassetto:2008yf},  providing matrix representations which are related to Yang-Mills theories in two dimensions and  successfully reproduce standard  perturbative techniques in flat space. Remarkably, some of these Wilson loops has been used in \cite{Correa:2012at, Bonini:2015fng} to derive a non-perturbative expressions for the generalized cusp anomalous dimension in the near-BPS limit which were tested by perturbation theory \cite{Correa:2012nk}, integrability \cite{Gromov:2012eu,Gromov:2013qga}, and AdS/CFT \cite{Drukker:2011za}.

In this paper, we address the problem of constructing and computing $\frac{1}{4}$-BPS supersymmetric Wilson loops in four-dimensional superconformal SU($N$) $\mathcal{N}=2$ super-Yang-Mills theories. This is primarily motivated by the AdS/CFT correspondence, where these observables can serve as alternative probes for testing holographic principles and various approaches in $\mathcal{N}=2$ theories 
\cite{Mitev:2015oty, Billo:2018oog, Galvagno:2021bbj, Giombi:2020mhz}
\footnote{See also \cite{Pilch:2000ue, Bobev:2013cja, Russo:2013qaa, Russo:2013kea, Buchel:2013id, Bobev:2018hbq, Russo:2019lgq} for another example of holography with reduced supersuymmetry.}.  

 In the rest of the Introduction, we will present these $\frac14$-BPS operators and summarize the main results of our analysis.

\subsection{$\frac{1}{4}$-BPS Wilson loops in $\mathcal{N}=2$ SYM}
The $\mathcal{N}=2$ vector multiplet consists of a gauge field $A_\mu$, two Weyl fermions $\lambda_I$, with $I=1,2,$ being the $su(2)$ R-symmetry index, and a complex scalar $\phi$ in the adjoint representation of SU$(N)$.
The supersymmetry variations of the bosonic fields are given by\footnote{The variations are those of \cite{Hama:2012bg}, up to set $\phi_{\textup{there}}=\frac{\mathrm{i}}{\sqrt{2}}\phi$, $\bar{\phi}_{\textup{there}}=\frac{\mathrm{i}}{\sqrt{2}}\bar{\phi}$. Some details on our conventions are collected in Appendix \ref{app2}.}
\begin{subequations}
	\label{eq:transformations}
	\begin{align}
		\delta A_\mu&=\mathrm{i}\xi^I\sigma_\mu\bar\lambda_I-\mathrm{i}\bar\xi^I\bar\sigma_\mu\lambda_I\,,\\[0.4em]
		\delta \phi&=-\sqrt{2}\xi^I\lambda_I \ , \\[0.4em]
		\delta\bar\phi&=\sqrt{2}\bar\xi^I\bar\lambda_I \,,
	\end{align}
\end{subequations} 
where $\xi^I$ is a conformal Killing spinor, while $\bar{\phi}$ is the complex conjugate of $\phi$.

  The general form of a supersymmetric Wilson loop incorporates the coupling with the complex scalar field and takes the following form\footnote{Note that eq. \eqref{eq:ansatz} can be generalized to an arbitrary representation by properly changing the trace and by normalizing the operator with the dimension of such representation.} 
\begin{equation}
	\label{eq:ansatz}
	W=
	\dfrac{1}{N}\tr\left[P\exp \int_\gamma dt\left(\mathrm{i}\dot x^\mu A_\mu +\mathrm{i}\frac{u\phi-v\bar\phi}{\sqrt{2}} \right)\right] \ , 
\end{equation}
where $P$ denotes the path-ordering symbol, while the curve $\gamma$,  along with the functions $u$ and $v$, fixes the amount of supersymmetry preserved by the operator. Applying the  variations (\ref{eq:transformations}) to (\ref{eq:ansatz}), we obtain the following BPS equations 
\begin{align}\label{eq:BPS}
	\bar{\xi}^I \bar{\sigma}_\mu\dot{x}^\mu-\mathrm{i}  u\xi^I=0\,, \qquad
	\xi^I \sigma^\mu\dot{x}^\mu+\mathrm{i}  v \bar{\xi}^I=0\,.
\end{align}
The existence of a local solution leads to the condition $v(x)u(x)=1$, which is satisfied if we choose  $u=\ee^{ -\mathrm{i} \varphi}$ and $ v=\ee^{\mathrm{i}\varphi}$, where $\varphi(x)$ can be interpreted  as an internal R-symmetry angle. When $\varphi=0$ and $\gamma$ is either a straight line or a circle, the operator  is  $\frac 12$-BPS \cite{Pestun:2007rz,Zarembo:2002an}. 

\begin{figure}[h]
	\begin{subfigure}{0.5\textwidth}
		\begin{center}
			\includegraphics[scale=0.5]{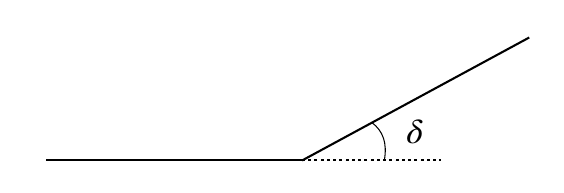}
		\end{center}
	\end{subfigure}
	\begin{subfigure}{0.5\textwidth}
		\begin{center}
			\begin{tikzpicture}[scale=0.4]
				\def\RadiusSphere{4} 
				\def\angEl{20} 
				\def\angAz{-20} 
				
				\shade[ball color = white!40, opacity = 0.5] (0,0) circle (\RadiusSphere);
				
				\pgfmathsetmacro\H{\RadiusSphere*cos(\angEl)} 
				\coordinate (O) at (0,0);
				\node[circle,draw,black,scale=0.3] at (0,0) {};
				\coordinate[mark coordinate] (N) at (0,\H);
				\coordinate[mark coordinate] (S) at (0,-\H);
				\draw[dashed, black](N)--(S);

				\tikzset{
					every path/.style={
						color=black
					}
				}
				\DrawLatitudeCircle[\RadiusSphere,dotted,thick]{0}
				%
				
				%

				\LongitudePlane[angle]{\angEl}{-80};
				\DrawLongitudeArc[black]{-50}{-90}{90}
				\path[angle] (00:\RadiusSphere) coordinate (Pprime);

				\LongitudePlane[angel]{\angEl}{-120};
				\DrawLongitudeArc[black]{-120}{-90}{90}
				\path[angel] (00:\RadiusSphere) coordinate (Oprime);

				\def\arcrad{2}
				\NewLatitudePlane[equator]{\RadiusSphere}{\angEl}{00};
				\draw[equator,-,black] (-120:\arcrad) arc (-120:-50:\arcrad);
				\path[equator] (-120:\arcrad) coordinate (m);
				\path[equator] (-90:\arcrad) coordinate (mprime);
				\draw[right] node[xshift=-0.3cm,yshift=-0.2cm] at (mprime){$\delta$};
				
				\draw[-,dashed] (Oprime) -- (O) -- (Pprime);

			\end{tikzpicture}
		\end{center}
	\end{subfigure}
	\caption{Graphical representations of the cusped Wilson loop in flat space on the left and of the wedge Wilson loop on $\mathbb{S}^2$ on the right.}\label{fig:cusp}
\end{figure}

To find less supersymmetric solutions to (\ref{eq:BPS}), we consider the ansatz 
	\begin{equation}
	\label{eq:cusp configuration}
	\begin{split}
		W&=
		\dfrac{1}{N} \tr 
		\left[ Pe^{\int_{-t_0}^0 dt\,L_1}Pe^{\int_0^{t_1} dt\,L_2}\right] \ ,
		\\[0.4em]
		L_1&= \mathrm{i}\dot x^\mu A_\mu +\mathrm{i}|\dot x|\frac{\phi-\bar\phi}{\sqrt{2}}\ , \\[0.4em]
		L_2&=\mathrm{i}\dot x^\mu A_\mu +\mathrm{i}|\dot x| \frac{e^{-\mathrm{i}\varphi}\phi-e^{\mathrm{i}\varphi}\bar\phi}{\sqrt{2}}\, ,
	\end{split}
\end{equation}
and apply  the transformations (\ref{eq:transformations}). A straightforward analysis leads to two  $\frac14$-BPS operators, i.e. the \emph{cusp} and the \emph{wedge}, represented in Fig.\ref{fig:cusp}.

The cusp configuration consists of two semi-infinite rays forming an angle of $\pi-\delta$ and is $\frac 14$-BPS when  $\varphi=\pm \delta$. Moreover, it reduces to the usual $\frac{1}{2}$-BPS straight line for $\delta\to 0$. Choosing $\varphi=\delta$, it is straightforward to verify that the conserved supercharges are $Q_{I2}+\bar Q_{I\dot2}$ and  $S_I^2-\bar S^{\dot2}_I$, where  $Q^I_\alpha$, $\bar Q_I^{\dot\alpha}$ are the familiar Poincar\'e supercharges, while  $S^I_\alpha$, $\bar S_I^{\dot\alpha}$ denote the superconformal charges. Altogether, they generate the full $4d$ $\mathcal{N}=2$ superalgebra (see Appendix \ref{app2} for more details). Notice that configurations with multiple Wilson rays in $\mathcal{N}=2$ theories appeared before in \cite{Kapustin:2006hi, Cordova:2016uwk, Gomez:2018usu}.
 
The wedge-like configuration is a special example of loop defined on $\mathbb{S}^2\subset\mathbb{R}^4$, consisting of two meridians separated by an opening angle $\delta$. In the limit $\delta \to \pi$, the wedge reduces to the circular configuration and undergoes a supersymmetry enhancement.

Since the wedge and  cusp configurations are related by a conformal transformation, also the wedge preserves four supercharges for $\varphi=\pm \delta$. Choosing $\varphi=\delta$, we derive the following expressions for the preserved charge   
\begin{equation}
	\mathcal{Q}_{+I}=\bar Q_{I\dot 2}+\mathrm{i}Q_{I2}+\left(\mathrm{i}S_I^2-\bar S_I^{\dot2}\right)\,, \qquad 
	\mathcal{Q}_{-I}=\bar Q_{I\dot 2}-\mathrm{i}Q_{I2}-\left(\mathrm{i}S_I^2+\bar S_I^{\dot2}\right)\,.
\end{equation}	
The resulting algebra is $\mathfrak{su}(1|2)\oplus\mathfrak{u}(1)$ 
\begin{equation}\label{eq:cuspalgebra}
	\begin{aligned}
		\acomm*{\mathcal{Q}_{+I}}{\mathcal{Q}^J_-}&= \left(\delta_I^J B-2{R^I}_J\right)\,,\\
		\comm*{B}{\mathcal{Q}_{+I}}&= \mathcal{Q}_{+I}\,, \qquad \comm*{B}{\mathcal{Q}_{-I}}=-\mathcal{Q}_{-I}\, ,
	\end{aligned}
\end{equation}
where ${R^I}_J$ are the $\mathfrak{su}(2)$ R-symmetry generators, while  $B$ and $F$ are spacetime generators 
\begin{equation}
	B= \mathrm{i}\left(P_{2\dot2}+K^{\dot22} \right) \,, \qquad
	F=\mathrm{i}\left(P_{1\dot1}+K^{\dot11} \right)\,.
\end{equation}
In the previous expressions, we denoted with $P_{\alpha\dot\beta}$ and $K^{\alpha\dot\beta}$  the translations and special conformal transformation generators, respectively.

\subsection{Main results}
The cusp and wedge configurations already appeared in $\mathcal{N}=4$ SYM \cite{Zarembo:2002an, Drukker:2007qr}. In this case, however,  the algebra preserved by these operators  is $\mathfrak{su}(1|2)\oplus\mathfrak{su}(1|2)$, where  the second R-symmetry $SU(2)$ factor arises from the freedom to  rotate the additional two complex scalars of the $\mathcal{N}=4$ vector multiplet which do not appear in the operator. Moreover, in \cite{Drukker:1999zq}, it was conjectured that the expectation value of the wedge in the maximally supersymmetric theory is related to that of the $\frac{1}{2}$-BPS circular loops by a rescaling of the coupling constant. Thus, at weak coupling and at  \emph{finite} $N$, it takes the form 
\begin{equation}
	\begin{split}
		\label{eq:vew1/4 N4 and its expansion}
	\big< W \big>_{\mathcal{N}=4}
		&= 1+ g^2\alpha^2\dfrac{C_F}{4}  + 	g^4\alpha^4\dfrac{C_F(2N^2-3)}{192} + g^6\alpha^6
		\dfrac{C_F(N^4-3N^2+3)}{4608N^2}+\ldots
	\end{split}
\end{equation} where we introduced the notation for \begin{equation}
\alpha^2=\dfrac{\delta(2\pi-\delta)}{\pi^2}  \ ,  \quad \quad \quad  C_F=\dfrac{(N^2-1)}{2N} \ .
\end{equation}  
The relation (\ref{eq:vew1/4 N4 and its expansion}) was explicitly tested by a direct calculation based on Feynman diagram techniques in \cite{Bassetto:2008yf} up to two loops. The result follows from a subtle cancellation between interacting diagrams suggesting a matrix model description for the observable and a corresponding localization procedure \cite{Pestun:2009nn}.

In contrast to the maximally supersymmetric case, the expectation value of the 
wedge configuration in $\mathcal{N}=2$ SYM theories  receives quantum corrections. We show below that in superconformal models with matter content in an arbitrary representation $\cR$ of SU($N$), the perturbative calculation takes the following form up to three loops
  \begin{equation}
	\label{eq:vev at three-loop def}
	\begin{split}
			\big<W\big> &= \big<W\big>_{\mathcal{N}=4} \Big |_{g^6} + 	 g^6\alpha^2\dfrac{C_F 3\zeta(3)\cC_4^\prime}{2^7\pi^4}  +\mathcal{O}(g^8) \\[0.4em]
			\cC_4^\prime& = \left(C_\cR i_\cR -\frac{N i_\cR}{2}-\frac{N^2}{2}\right) \ ,
	\end{split}
\end{equation} where $i_\cR$ and $C_\cR$ denote, respectively, the Dynkin index and quadratic Casimir\footnote{The Dynkin index is defined by the relation $\Tr_{\cR} = T^a T^b =i_\cR \delta^{ab}$, while $C_\cR = T^a T^a$, with $a,b =1,\ldots N^2-1$. } of the representation $\cR$. The previous expression implies that the wedge actually coincides with its $\mathcal{N}=4$ counterpart up to two loops. The deviation from the maximally supersymmetric case occurs at three loops and involves the colour coefficient $\mathcal{C}_4^\prime$. As it will be clear in the following, this result follows from the conformality consideration. Let us also stress that in the limit $\delta=\pi$, the observable coincides with the vacuum expectation of the $\frac{1}{2}$-BPS circular Wilson loop derived in \cite{Billo:2019fbi,Andree:2010na}, as expected. 

The symmetry properties of the wedge operator pose the question of whether its expectation value  (\ref{eq:vev at three-loop def}) can be described by a  localization procedure and a corresponding matrix model. 	To do so, a  localizing supercharge which annihilates the operator is required.  Naively, it is tempting to consider the charge  employed to localize the wedge in $\mathcal{N}=4$ SYM \cite{Pestun:2009nn}. However, this supercharge  belongs to the diagonal sum of the algebra $\mathfrak{su}(1|2)\oplus\mathfrak{su}(1|2)$ which, as we previously mentioned,  is only available in the maximally supersymmetric theory.

Although a genuine localization approach for the wedge in $\mathcal{N}=2$ SYM does not seem applicable with the current results of the literature, we are able to construct a representation for the observable based on the matrix model generated by supersymmetric localization on the four-sphere \cite{Pestun:2007rz}. We show in Section \ref{sec4} that 
\begin{align}
	\label{eq:v.e.v matrix model intro}
	\big<W\big> &=\dfrac{1}{\cZ} \int \dd a \ \ee^{-\tr a^2-S_{\rm int}(a,g)} \dfrac{1}{N} \tr \exp(a\dfrac{\alpha g}{\sqrt{2}})\\[0.8em]
		{S}_{\rm int}(a,g) &=-  \sum_{p=2}^{\infty}\left(-\dfrac{g^2}{8\pi^2}\right)^{p}\frac{\zeta(2p-1)}{p}\Tr_\cR^\prime a^{2p}\ ,
	\end{align} perfectly reproduces (\ref{eq:vev at three-loop def}). In the previous expression, $\cZ$ is given by (\ref{eq:v.e.v matrix model intro})  without the insertion of the phase factor, $a$ is an $su(N)$ matrix describing the zero modes of the vector-multiplet scalar $\phi$,  while $\Tr_{\cR}^\prime =\Tr_{\cR}-\Tr_{\rm Adj}$. 
For a special class of superconformal theories, consisting of hypermultiplets in the rank-two symmetric and antisymmetric representation of SU($N$), the coefficient $\cC_{4}^\prime$ in (\ref{eq:vev at three-loop def}) vanishes and it is possible to calculate the observable at four loops due to drastic simplification of the different Feynman diagrams (see Appendix \ref{sec:four-loop corrections in E theory}). Also in this case, the matrix integral (\ref{eq:v.e.v matrix model intro}) reproduces the perturbative results.  

As a consequence of this non-trivial analysis, it is tempting to conjecture that \eqref{eq:v.e.v matrix model intro} could actually provide a non-perturbative expression for the observable which allows us to study the strong coupling limits in different theories. 
Since the wedge is an extended object, we expect that it can encode non-trivial information about dual holographic descriptions and consequently, it is interesting to apply our formalism (\ref{eq:v.e.v matrix model intro}) to a model for which the strong coupling regimes and the holographic dual descriptions remain largely mysterious, i.e. conformal $\mathcal{N}=2$ SQCD \cite{Gadde:2009dj, Gaiotto:2009gz, OColgain:2011kej, Reid-Edwards:2010vpm, Dei:2024frl}. Using the formalism developed in \cite{Passerini:2011fe}, we show that, in the large-$N$ limit and at strong coupling, the observable exhibits different behaviors depending on the value of the parameter $\alpha$.
If $\alpha>1/4$, the Wilson loop follows a power law behavior analogous to the circular case \cite{Passerini:2011fe}. 
\begin{equation}
	\big< W\big>_{\rm SQCD} \stackrel{\lambda \to \infty}{\simeq}
\frac{R_\alpha}{C}\left(\frac{\pi}{2\log\left(\frac{\lambda}{C}\right)}\right)^{2(\alpha-1/4)}\left(\frac{\lambda}{C}\right)^{4(\alpha-1/4)} \ , \quad \mathrm{if}\quad \alpha>1/4 \ ,
\end{equation} 
where $C$ and $R_\alpha$ are $\lambda$-independent constants given in eq. \eqref{eq:Cdef} and \eqref{eq:Ra def}, respectively.
However, when $\alpha<1/4$, the expectation value approaches a $\lambda$-independent value
\begin{equation}
	\big< W\big>_{\rm SQCD} \stackrel{\lambda \to \infty}{\simeq}\frac{1}{\cos 2\pi \alpha} \ ,  \quad \mathrm{if}\quad \alpha<1/4 \ .
\end{equation} 
This discontinuous behavior in $\alpha$ or, equivalent,y in the opening angle $\delta$, is a new phenomenon for the wedge, which drastically differs from its $\mathcal{N}=4$ counterpart.

The remaining sections of this paper are organized as follows.  Section \ref{sec3} is devoted to a detailed perturbative computation in flat space.  In Section \ref{sec4},  we show that the matrix integral (\ref{eq:v.e.v matrix model intro}) for the observable reproduces the perturbative results.  The analysis of  ${\cal N} =2 $ superconformal QCD is examined in Section \ref{sec5}, where the large 't Hooft coupling limit is thoroughly discussed. Our conclusions and outlooks are contained in Section \ref{sec7}. This work is supplemented with two appendices. In the first one we discuss the supersymmetry algebra of the wedge configuration, while in the second one, we test our matrix model representation for the observable at \emph{four loops} in a particular superconformal model, known as the $\mathbf{E}$ theory.

\section{The $\frac{1}{4}$-BPS Wilson loop in flat space}\label{sec3}
\label{sec:1/4 flat space}
In this section, we study  the expectation value of the  $\frac{1}{4}$-BPS wedge Wilson loop  in perturbation theory. We will consider superconformal SU($N$) $\mathcal{N}=2$ SYM theories massless hypermultiplets in a generic representation $\cR$. Conformal invariance implies that the (one-loop exact)  $\beta$-function \cite{Billo:2019job, Jones:1974mm, Howe:1984xq,Billo:2024fst}  \begin{equation}
		\label{eq:beta-function}
		\beta(g)= \dfrac{(i_\cR-N)}{8\pi^2} g^3 \ ,
\end{equation} vanishes. In the previous expression, $i_\cR$ is defined via the relation  $\Tr_\cR T^a T^b=\delta^{ab}i_\cR$, and satisfies the condition  $i_\cR=N$.

 According to (\ref{eq:cusp configuration}), the wedge operator in flat space  takes the following
   \begin{align}
	\label{eq:1/4 Wilson loop}
	W_{\frac{1}{4}}&=\dfrac{1}{N}\tr\left( P\ee^{\int_{-\infty}^0 dt\,L_1}P\ee^{\int_0^{\infty} dt\,L_2}\right)\,,\\[0.4em]
L_1&= g\left(\mathrm{i}\dot x^\mu A_\mu -\dfrac{\mathrm{i }}{\sqrt{2}} \left(\phi-\bar\phi\right)\right)\, , \notag \\[0.4em]
L_2&=g\left(\mathrm{i}\dot x^\mu A_\mu +\dfrac{\mathrm{i }}{\sqrt{2}}\left(e^{-\mathrm{i}\delta}\phi-e^{\mathrm{i}\delta}\bar\phi\right)\right)\notag \, .
\end{align} Mapping the Wilson loop contour on $\mathbb{S}^2$ to the flat space by the stereographic projection, we obtain the following parametrization of the coordinates $x^\mu(t)$  \begin{equation}
	\label{eq:parametrization}
	x^\mu(t) = \begin{cases}
		\left(-\dfrac{2t}{1+t^2},\ 0, \ \dfrac{1-t^2}{t^2+1}\right) \ ,  \quad \quad \quad \quad \quad \quad \quad \ \  \  -\infty< t \le 0 \ ,\\[1em]
		\left(\dfrac{2t}{1+t^2}\cos\delta,\ \dfrac{2t}{1+t^2}\sin\delta, \ \dfrac{1-t^2}{t^2+1}\right) \ , \ \ \ \ \  \quad   \quad 0\le t < \infty\ .
	\end{cases}
\end{equation}   

 In perturbation theory, we expand the vacuum expectation value of (\ref{eq:1/4 Wilson loop}) in a power series of the gauge coupling constant as follows \begin{equation}
	\label{eq:pert expansion}
	\big<W\big> = \sum_{k=0}^{\infty} g^{2k} \mathcal{W}_{2k}  \ ,
\end{equation} where  $\mathcal{W}_0=1$. The quantities $\cW_{2k}$, for $k\ne0$, carry the dependence of the observable on the matter representation  $\cR$ and are finite in four dimensions as a  consequence of conformal invariance.

\paragraph{Organizing perturbation theory}
An effective approach to organize perturbative calculations of observables in common with $\mathcal{N}=4$ SYM is the so-called  \emph{difference theory method} \cite{Billo:2019fbi,Andree:2010na}. It is based on the observation that at any perturbative order $ g^{2k}$,  we can decompose the quantities $\cW_{2k}$ of eq.\1(\ref{eq:pert expansion}) as follows: 
\begin{equation}
	\label{W2kis}
	\cW_{2k} = \cW_{2k}^{\rm m.e.} + \cW_{2k}^{\rm v.m.} + \cW_{2k}^{\cR}~. 
\end{equation}	
We denoted with $\cW_{2k}^{\cR}$ the diagrams involving internal \emph{matter} lines  in the representation $\cR$, while the first two contributions capture, respectively,  multiple-exchange diagrams, in which the gauge field $A_\mu$ and the scalar field $\phi$ are exchanged at tree-level, and the interaction corrections involving vector-multiplet lines  only. This means that $\cW_{2k}^{\rm v.m.}$ and $ \cW_{2k}^{\rm m.e.}$ are in common with the $\cN=4$ theory and can be employed to simplify the calculation. To see this, we observe that for $\cR=\rm Adj$, we precisely obtain the perturbative expansion of the observable in $\mathcal{N}=4$ SYM, i.e. \begin{equation}
		\label{W2kis in N4}
	\cW^{\mathcal{N}=4}_{2k} = \cW_{2k}^{\rm m.e.} + \cW_{2k}^{\rm v.m.} + \cW_{2k}^{\rm Adj}~ .
\end{equation} Since the  left-hand side of the previous expression is given by  eq. (\ref{eq:vew1/4 N4 and its expansion}), we can combine together eq.s (\ref{W2kis}) and (\ref{W2kis in N4}) to finally obtain \footnote{Eq. (\ref{W2kis}) is strictly  valid in superconformal models. In fact, the right-hand side of eq. (\ref{W2kis in N4}) precisely reproduces the expansion of eq. (\ref{eq:vew1/4 N4 and its expansion}) after having regularized and removed divergent intermediate contributions.  This operation usually holds up to  $\cO(d-4)^n$ terms, with $n$ being a positive integer.  These evanescent contributions are completely irrelevant in superconformal models but play an important role in theories with non-vanishing $\beta$-function (see Section 3 of \cite{Billo:2024fst} for a detailed example on this point regarding circular Wilson loop). } \begin{equation}
\label{W2kis def}
\cW_{2k}=\cW_{2k}^{\mathcal{N}=4} + \cW_{2k}^\prime \ , \quad \text{where} \quad  \cW_{2k}^\prime \equiv \cW_{2k}^{\cR} - \cW_{2k}^{\rm Adj} \ .
\end{equation}  In conclusion, at  any perturbative order, the  result is given by that in $\mathcal{N}=4$ SYM with an additional contribution, i.e.  $\cW_{2k}^\prime $, which is obtained  by subtracting from $\cW_{2n}^\cR$ the same diagrams in which the internal lines are  in the adjoint representation.  These contributions are sometimes denoted in the literature \cite{Andree:2010na,Billo:2019fbi,Billo:2024fst} as  \textit{difference theory} terms. 

\subsection{One-loop corrections}
\label{sec:one-loop corrections}
 At one-loop accuracy, the difference theory diagram  $\cW_2^\prime$ (\ref{W2kis def}), vanishes since all the diagrams arise from the propagation of vector-multiplet fields. As a result, the observable coincides  with its  $\mathcal{N}=4$ counterpart. However, for future convenience, we present explicit structure of the one-loop diagrams. 

 Let us begin with recalling that in the Feynman gauge, the tree-level propagator of the complex scalar $\phi$ and the gauge field $A_\mu$ are identical up to spacetime indices, i.e. 
\begin{equation}
	\big<A^a_\mu(x)A^b_\nu(y)\big> =\delta_{\mu \nu}\delta^{ab}\Delta(x_{12})\ , \quad \ 	\big<\phi^a(x)\Bar{\phi}^b(y)\big> =\delta^{ab} \Delta(x_{12}) \ .
\end{equation} In the previous expression, $a,b=1,\ldots,N^2-1$ are colour indices, we introduced the notation  $x_{12}\equiv x_1-x_2$, while $\Delta(x_{12})$ is the usual massless propagator in $d$ dimensions: \begin{align}
	\label{eq:tree-level prop}
	\Delta(x_{12}) &= \dfrac{\Gamma(d/2-1)}{4\pi^{d/2}(x^2_{12})^{d/2-1} } =\dfrac{f(d)}{(x^2_{12})^{d/2-1} }
\ .
\end{align}

At order $g^2$,  we encounter two classes of \textit{single-exchange} diagrams which were originally evaluated in  \cite{Bassetto:2008yf}. Going through the calculation, we find  that 
\begin{align}
	\label{eq:one-loop single exchange}
	\mathord{ \begin{tikzpicture}[baseline=-0.65ex,scale=0.6]
		\draw
		(-2,-1) coordinate (a) 
		-- (0,2) coordinate (b) 
		-- (2,-1) coordinate (c) pic["$\delta$",  draw, angle eccentricity=1.2, angle radius=1cm]
		{angle=a--b--c}; 
		\begin{feynman}
			\vertex (C) at (-1.65,-.5);
			\vertex (D) at (1.65,-.5);
			\diagram*{
				(C) -- [photon] (D),
				(C) -- [photon, fermion,thick] (D),
			};
		\end{feynman}
	\end{tikzpicture}
} + 	\mathord{ \begin{tikzpicture}[baseline=-0.65ex,scale=0.6]
			\draw
			(-2,-1) coordinate (a) 
			-- (0,2) coordinate (b) 
			-- (2,-1) coordinate (c) pic["$\delta$",  draw, angle eccentricity=1.2, angle radius=1cm]
			{angle=a--b--c}; 
			\begin{feynman}
				\vertex (C) at (0.85,0.7);
				\vertex (D) at (2,-1);
				\diagram*{
					(C) -- [photon, half left] (D),
					(C) -- [anti fermion, half left] (D),
				};
			\end{feynman}
		\end{tikzpicture}
	} 
	=g^2C_Ff(d)(2\pi-\delta)\delta \ ,
\end{align}  where $C_F=(N^2-1)/2N$ is the fundamental Casimir. In the previous expression,  we employed a double continuos/wiggly line notation to represent, respectively,  the  scalar and gauge-field propagator. The form of eq. (\ref{eq:one-loop single exchange}) is  general and can be employed to calculate  single-exchange diagrams dressed with arbitrary loop corrections to the  propagators, provided that these are finite in the limit $d\to 4$. Finally, using  eq. (\ref{eq:tree-level prop}), we  find that 
 \begin{align}
	\label{eq:net result w22}
	g^2\cW_2=g^2\alpha^2\dfrac{C_F}{4} \ ,  
\end{align}
where we recall that $\alpha$ is defined in eq. (\ref{eq:vew1/4 N4 and its expansion}). As anticipated, the previous expression is in agreement with the results derived in \cite{Drukker:2007qr,Bassetto:2008yf} for $\mathcal{N}=4$ SYM.

	\subsection{Two-loop corrections}
	\label{sec:two-loop corrections 1/4}
	It turns out that  the observable coincides with the  $\mathcal{N}=4$ counterpart also at two-loop accuracy. To see this, it is sufficient to observe that the only difference between the two models involves diagrams with one-loop corrections to the propagators. Since these are UV divergent in $d=4$, the conformality condition implies that these contributions vanish. 
	
To show this explicitly, we apply eq. (\ref{W2kis in N4}) to  write the two-loop correction as follows 
	\begin{equation}
		\begin{split}
			\label{eq:two-loop correction 1/4}
			g^4\cW_4 =
			g^4\alpha^4\dfrac{C_F(2N^2-3)}{192}
			+ g^4\cW^\prime_4 \ . 
		\end{split}
	\end{equation} The difference-theory term $g^4\cW^\prime_4$ arises from diagrams involving matter lines in the representation $\cR$ to which we subtract the same contribution with $\cR=\rm Adj$. Diagrammatically, we can represent these terms as follows 
\begin{equation}
	\label{eq:delta4}
	g^4	\cW^\prime_4 =  	\mathord{ \begin{tikzpicture}[baseline=-0.65ex,scale=0.6]
			\draw
			(-3,-2) coordinate (a) 
			-- (0,2) coordinate (b) 
			-- (3,-2) coordinate (c) pic["$\delta$",  draw, angle eccentricity=1.2, angle radius=1cm]
			{angle=a--b--c}; 
			\begin{feynman}
				\vertex (C) at (1.2,0.4);
				\vertex (D) at (2.9,-1.8);
				\newcommand\tmpda{0.4cm}
				\newcommand\tmpdb{2.cm}
				\diagram*{
					(C) -- [photon, half left] (D),
					(C) -- [plain, half left, with arrow=\tmpdb] (D),
					(C) -- [plain, half left, with arrow=\tmpda] (D),
				};
			\end{feynman}
			\filldraw[color=white!80, fill=white!15](3,0) circle (0.9);	
			\draw [black] (3,0) circle [radius=0.9cm];
			\draw [black,dashed,thick] (3,0) circle [radius=0.8cm];
		\end{tikzpicture}
	}  \ + \  	\mathord{ \begin{tikzpicture}[baseline=-0.65ex,scale=0.6]
			\draw
			(-3,-2) coordinate (a) 
			-- (0,2) coordinate (b) 
			-- (3,-2) coordinate (c) pic["$\delta$",  draw, angle eccentricity=1.2, angle radius=1cm]
			{angle=a--b--c}; 
			\begin{feynman}
				\vertex (C) at (-2.6,-1.5);
				\vertex (D) at (2.6,-1.5);
				\newcommand\tmpda{0.4cm}
				\newcommand\tmpdb{2.6cm}
				\diagram*{
					(C) -- [photon] (D),
					(C) -- [plain,with arrow=\tmpda] (D),
					(C) -- [plain,with arrow=\tmpdb] (D),
				};
			\end{feynman}
			\filldraw[color=white!80, fill=white!15](0,-1.4) circle (0.9);	
			\draw [black] (0,-1.4) circle [radius=0.9cm];
			\draw [black,dashed,thick] (0,-1.4) circle [radius=0.8cm];
		\end{tikzpicture}
	} \ . 
\end{equation} 
The internal self-energies correspond to the one-loop correction to the propagators in the \textit{difference theory} approach. These diagrams are evaluated explicitly in Appendix of  \cite{Billo:2024fst} (see in particular eq.s\1(B.13,B.14)) where it is showed that 
	\begin{equation}
	\label{eq:prime trace}
		\begin{split}
		\mathord{
			\begin{tikzpicture}[scale=0.6, baseline=-0.65ex]
				\filldraw[color=white!80, fill=white!15](0,0) circle (1);	
				\draw [black] (0,0) circle [radius=1cm];
				\draw [black, thick, dashed] (0,0) circle [radius=0.9cm];
				\begin{feynman}
					\vertex (A) at (-2,0);
					\vertex (C) at (-1,0);
					\vertex (B) at (1, 0);
					\vertex (D) at (2, 0);
					\diagram*{
						(A) -- [fermion] (C),
						(B) --[fermion] (D),
						(A) -- [photon] (C),
						(B) --[photon] (D),
					};
				\end{feynman}
			\end{tikzpicture} 
		}
	 \propto \delta^{ab}(i_\cR-N)\ .
\end{split} 
\end{equation}
  Since in superconformal models we have $i_\cR=N$, eq.\1(\ref{eq:prime trace})  vanishes and the final result  
coincides with that of $\mathcal{N}=4$ SYM. 

	\subsection{Three-loop corrections}
	\label{sec:three-loop corrections}
	In the previous subsection, we showed that the one- and two-loop corrections to  the vacuum expectation value of the wedge  (\ref{eq:1/4 Wilson loop}) coincides with the result in $\mathcal{N}=4$ SYM. At three loops, this property ceases to hold for an arbitrary representation $\cR$.
	
	According  to eq. (\ref{W2kis def}), we can express the three-loop corrections  as follows  \begin{equation}
		\label{eq:three-loop correction}
		g^6\cW_6=g^6\alpha^6\dfrac{C_F}{6!8}\dfrac{5(N^4-3N^2+3)}{4N^2} + g^6\cW^\prime_6  \ ,
	\end{equation}  where the first term on the right-hand side is  the three-loop contribution to the observable in  $\mathcal{N}=4$ SYM  (\ref{eq:vew1/4 N4 and its expansion}).
 The difference-theory term  $g^6\cW^\prime_6$ is obtained by subtracting from the three-loop interaction diagrams with internal lines in the representation $\cR$ analogous contributions  with $\cR=\rm Adj$.  These corrections can be organized in terms of three classes of diagrams according to the number of insertions on the cusp rays. We use the notation \begin{equation}
	\label{eq:notation}
	\cW_6^\prime = \cW_{6(2)}^\prime+\cW_{6(3)}^\prime+\cW_{6(4)}^\prime \ ,
\end{equation} to distinguish each contribution. We show below that only $\cW_{6(2)}^\prime$ provides a non-trivial contribution.
	\paragraph{Diagrams with two insertions} 
The diagrams characterized by two insertions on the operator contour take the following form   \begin{equation}
		\label{eq:delta6 bubble}
		g^6	\cW^{\prime}_{6(2)} =  	\mathord{ \begin{tikzpicture}[baseline=-0.65ex,scale=0.6]
				\draw
				(-3,-2) coordinate (a) 
				-- (0,2) coordinate (b) 
				-- (3,-2) coordinate (c) pic["$\delta$",  draw, angle eccentricity=1.2, angle radius=1cm]
				{angle=a--b--c}; 
				\begin{feynman}
					\vertex (C) at (1.2,0.4);
					\vertex (D) at (2.9,-1.8);
					\newcommand\tmpda{0.4cm}
					\newcommand\tmpdb{2.cm}
					\diagram*{
						(C) -- [photon, half left] (D),
						(C) -- [plain, half left, with arrow=\tmpdb] (D),
						(C) -- [plain, half left, with arrow=\tmpda] (D),
					};
				\end{feynman}
				\filldraw[color=gray!80, fill=gray!15](3,0) circle (0.9);	
				\draw [black] (3,0) circle [radius=0.9cm];
				\draw [black,dashed,thick] (3,0) circle [radius=0.8cm];
				\begin{feynman}
					\vertex (c)  at (3.1,0) {\text{\footnotesize 2-loop }\normalsize} ;
				\end{feynman}
			\end{tikzpicture}
		}  \ + \  	\mathord{ \begin{tikzpicture}[baseline=-0.65ex,scale=0.6]
				\draw
				(-3,-2) coordinate (a) 
				-- (0,2) coordinate (b) 
				-- (3,-2) coordinate (c) pic["$\delta$",  draw, angle eccentricity=1.2, angle radius=1cm]
				{angle=a--b--c}; 
				\begin{feynman}
					\vertex (C) at (-2.6,-1.5);
					\vertex (D) at (2.6,-1.5);
					\newcommand\tmpda{0.4cm}
					\newcommand\tmpdb{2.6cm}
					\diagram*{
						(C) -- [photon] (D),
						(C) -- [plain,with arrow=\tmpda] (D),
						(C) -- [plain,with arrow=\tmpdb] (D),
					};
				\end{feynman}
				\filldraw[color=gray!80, fill=gray!15](0,-1.4) circle (0.9);	
				\draw [black] (0,-1.4) circle [radius=0.9cm];
				\draw [black,dashed,thick] (0,-1.4) circle [radius=0.8cm];
				\begin{feynman}
					\vertex (c)  at (0.1,-1.4) {\text{\footnotesize 2-loop }\normalsize} ;
				\end{feynman}
			\end{tikzpicture}
		}  \ , 
	\end{equation} where the internal double dashed/continuos  bubble  denotes  the two-loop corrections to the adjoint scalar and gauge field propagator in the difference theory approach. For an arbitrary representation $\cR$, these corrections are given by eq.s\1(B.33,B.34) in Appendix B.2 of \cite{Billo:2024fst}. In particular, using these relations, we find that 
 \begin{align}
		\label{eq:2-loop diagrammatic corrections difference loop scalar}
		\mathord{
			\begin{tikzpicture}[scale=0.55, baseline=-0.65ex]
				\filldraw[color=gray!80, fill=gray!15](0,0) circle (1);	
				\draw [black] (0,0) circle [radius=1cm];
				\draw [black, thick, dashed] (0,0) circle [radius=0.9cm];
				\begin{feynman}
					\vertex (A) at (-2,0);
					\vertex (C) at (-1,0);
					\vertex (C1) at (0.1,0) {\text{\footnotesize 2-loop }\normalsize} ;
					\vertex (B) at (1, 0);
					\vertex (D) at (2, 0);
					\diagram*{
						(A) -- [fermion] (C),
						(B) --[fermion] (D),
					};
				\end{feynman}
			\end{tikzpicture} 
		}&=g^4\delta^{ab} \Delta^{(2)}(x_{12})
 \  ,  \\[0.6em]
	\label{eq:2-loop diagrammatic corrections difference loop gauge}
	\mathord{
		\begin{tikzpicture}[scale=0.55, baseline=-0.65ex]
			\filldraw[color=gray!80, fill=gray!15](0,0) circle (1);	
			\draw [black] (0,0) circle [radius=1cm];
			\draw [black, thick, dashed] (0,0) circle [radius=0.9cm];
			\begin{feynman}
				\vertex (A) at (-2,0);
				\vertex (C) at (-1,0);
				\vertex (C1) at (0.1,0) {\text{\footnotesize 2-loop }\normalsize} ;
				\vertex (B) at (1, 0);
				\vertex (D) at (2, 0);
				\diagram*{
					(A) -- [photon] (C),
					(B) --[photon] (D),
				};
			\end{feynman}
		\end{tikzpicture} 
	} 
&=g^4\delta^{ab}\left(\delta_{\mu \nu}\Delta^{(2)}(x_{12})+\partial_{1,\mu}\partial_{1,\nu}\Delta^{(2),\rm g}(x_{12})\right) \notag  \ .
	\end{align} In the previous expression, the one-loop correction to the adjoint scalar field in the difference method, i.e.  $\Delta^{(2)}(x_{12})$, is explicitly given by  \begin{equation}
	\label{eq:Delta2}
 \Delta^{(2)}(x_{12}) =	 \dfrac{ \cC_{4}^\prime 6\zeta(3)\Gamma(3d/2-5)}{4^{8-d}\pi^{4+d/2}\Gamma(5-d)(x^2_{12})^{3d/2-5}} \equiv \dfrac{f^{(2)}(d)}{(x_{12}^2)^{3d/2-5}} \ ,
\end{equation} where the colour factor  $\cC_4^\prime$ is defined in (\ref{eq:vev at three-loop def}).
 By a direct calculation, it is straightforward to verify the gauge-like term $\partial_{1,\mu}\partial_{1,\nu}\Delta^{(2),\rm g}(x_{12})$ does not contribute within the diagrams of eq.\1(\ref{eq:delta6 bubble}), as expected by gauge invariance. 
	
Since  eq.\1(\ref{eq:Delta2}) is completely finite for $d\to4 $,   we determine the single-exchange diagrams (\ref{eq:delta6 bubble}) by employing eq.\1(\ref{eq:net result w22}) with the replacement $f(d)\to f^{(2)}(d)$, i.e.   
		\begin{equation}
		g^6\cW_6^{(1)}  = g^2 C_F f^{(2)}(d)\pi^2 \alpha^2
		\Big|_{d=4}=g^6\alpha^2\dfrac{C_F 3\zeta(3)\cC_4^\prime}{2^7\pi^4}
\ , 
		\end{equation} where we recall that the parameter $\alpha$ is defined in eq.\1(\ref{eq:vew1/4 N4 and its expansion}). As we already anticipated, the three-loop correction differs from that of $\mathcal{N}=4$ SYM for an arbitray representation $\cR$. Moreover, let us also note that at this perturbative order the parameter $\alpha$ does not appear with the same power of the gauge coupling constant $g$. 
		
		\paragraph{Absence of other contributions}
		Let us finally discuss the  terms   $\cW_{6(3)}^\prime$ and $\cW_{6(4)}^\prime$ we introduced in  (\ref{eq:notation}). These contributions are associated with difference-theory diagrams with three and four emissions from the Wilson loop contour, respectively. 
		
		First, we observe that  $\cW_{6(4)}^\prime$ vanishes trivially. In fact, the only three-loop diagrams we can construct with four emissions from the contour and with internal lines in the representation $\cR$ necessarily  involves the one-loop self-energies eq.\1(\ref{eq:prime trace}). Since these propagators vanish in superconformal setups, $\cW^\prime_{6(4)}$ does not contribute to the observable.
		
		It remains to show that  $\cW^\prime_{6(3)}$ vanishes as well. This  class of contributions involves  diagrams with three emissions from the Wilson loop contour and internal lines in the representation $\cR$. At order $g^6$, these contributions arises  from  the \textit{irreducible} one-loop corrections in the difference method to the pure- and scalar-gauge vertex, i.e. \begin{equation}
			\label{eq:bigspider}
			\begin{split}
				\mathord{
					\begin{tikzpicture}[radius=2.cm, baseline=-0.65ex, scale=0.6]
						\draw [black] (0,0) circle [radius=0.8cm];
						\draw [black, dashed] (0,0) circle [radius=0.7cm];
						\begin{feynman}
							\vertex (A) at (0,2);
							\vertex (C) at (0,0.8);
							\vertex (D) at (-1.5, -1.3);
							\vertex (B) at (-0.7, -0.4);
							\vertex (B1) at (0.7,-0.4);
							\vertex (B2) at (1.5,-1.3);
							\diagram*{
								(A) -- [photon] (C),
								(B) --[ anti charged scalar] (D),
								(B) --[photon] (D),
								(B1) --[ charged scalar] (B2),
								(B1) --[photon] (B2),
							};
						\end{feynman}
					\end{tikzpicture} 
				} \  .
			\end{split}
		\end{equation}  A detailed calculation of these diagrams is presented in Appendix E of \cite{Billo:2024fst}, where the authors show that they are proportional to 
	 \begin{equation}
			\label{eq:colour factro big spider two adjoint scalars}
			\begin{split}
				 \left(\Tr^\prime_\cR T^a T^b T^c + \Tr^\prime_{\bar{\cR}} T^a T^b T^c\right) \propto f^{abc} (i_\cR-N)\ .
			\end{split}
		\end{equation} Since superconformal models have $i_\cR=N$, the quantity $\cW^\prime_{6(3)}$ does not contribute  to the observable. As a result, we find that the three-loop correction deviates from the result in $\mathcal{N}=4$ SYM by a term proportional to $\zeta(3)$, i.e.
\begin{equation}
	\label{eq:three-loop correction def}
	g^6\cW_6=g^6\alpha^6\dfrac{C_F}{6!8}\dfrac{5(N^4-3N^2+3)}{4N^2} + g^6\dfrac{C_F 3\zeta(3)\cC_4^\prime}{2^7\pi^4}\alpha^2 \ ,
\end{equation} Since the one- and two-loop corrections coincide with the $\mathcal{N}=4$ counterparts, we explicitly verified  (\ref{eq:vev at three-loop def}).

	\section{The matrix model on $\mathbb{S}^4$}\label{sec4}
	\label{sec:the matrix model}
	In this section, we study the $\frac{1}{4}$-BPS wedge Wilson loop on the four-sphere $\mathbb{S}^4$. We propose an explicit expression for the expectation value of the operator based on the matrix model generated by supersymmetric localization \cite{Pestun:2007rz} for $\mathcal{N}=2$ SYM theories and show that  it matches the result presented in the previous section. 
 	\subsection{The partition function on $\mathbb{S}^4$}
	Let us consider a general SU($N$) $\mathcal{N}=2$ SYM theory defined on a unit\footnote{The dependence on the radius $r$ can be restored in the integrand of eq. (\ref{eq:partition function}) by the replacement  $a \to ra$.} four-sphere $\mathbb{S}^4$. Supersymmetric localization maps the partition function to an interacting matrix model 
	\cite{Pestun:2007rz}   
	\begin{equation}
		\label{eq:partition function}
		\mathcal{Z} = \int \dd a  \ \mathrm{e}^{-\frac{8\pi^2 }{{g}^2}\tr a^2} Z^\mathcal{R}_{\text{1-loop}}  \left|Z_{\rm Inst}\right|^2 \ .
	\end{equation} 
In the previous expression, ${g}$ is the coupling constant\footnote{In conformal setups, where the $\beta$-function vanishes, $g$ is a pure dimensionless quantity. On the other hand, when the $\beta$-function is non-vanishing, $g$ runs at the quantum level and has to be interpreted as the running coupling constant evaluated at the scale $E=1/r$, see \cite{Pestun:2007rz,Billo:2023igr,Billo:2024hvf} for more details.}, while  $a$ is  a traceless hermitian $N\times N$ matrix describing the zero modes of the vector-multiplet  scalar field on $\mathbb{S}^4$. Decomposing the matrix $a$ in terms of the hermitian generators $T^a$ of $su(N)$, i.e. $a=a_bT^b$ with $b=1,\ldots,N^2-1$, the    integration measure $\dd a  $ (\ref{eq:partition function}) is given by  \begin{equation}
		\label{eq:flat integration measure matrix model}
	\dd  a=\prod_{b=1}^{N^2-1}\dd a_b \ , \quad \quad \text{with}  \quad \quad   \tr T_a T_b =\dfrac{\delta_{ab}}{2}~ . 
	\end{equation}
In addition to the Guassian term, the partition function on the $\mathbb{S}^4$ (\ref{eq:partition function}), also involves the instanton contributions $Z_{\rm Inst}$, which we can ignore since we will work in perturbation theory, and the contributions of the one-loop fluctuation determinants \cite{Pestun:2009nn}
   \begin{equation}
			\label{eq:one-loop in general superconformal theories}
			\left|Z^\mathcal{R}_{\text{1-loop}}\right|^2 = \dfrac{\prod_{\mathbf{w}_{\rm Adj}} H( \mathrm{i}\mathbf{w}_{\rm Adj}\cdot \mathbf{a})}{\prod_{\mathbf{w}_{\mathcal{R}}} H( \mathrm{i} \mathbf{w}_\mathcal{R}\cdot \mathbf{a})} \ .
	\end{equation} In the previous expression, $\mathbf{a}$ is a $N$-dimensional vector containing the real eigenvalues $a_u$ of the matrix $a$, while $\mathbf{w}_\mathcal{R}$ and $\mathbf{w}_{\rm Adj}$ are the weight vectors of the representation $\mathcal{R}$ and of the adjoint one, respectively. In eq. (\ref{eq:one-loop in general superconformal theories}), $H(x)$ is given by 
 \begin{equation}
 	\label{H}
	\begin{split}
	H(x)
	&=G(1+x)G(1-x)\\[0.4em]
	&= \exp( -x^2(1+\gamma_E)-\sum_{p=1}^{\infty}\frac{\zeta(2p+1)}{p+1}x^{2p+2} )   \ ,
\end{split}
\end{equation} where $G(x)$ is the Barnes $G$-function and  the second relation holds for small $x$. 

 Following the approach of \cite{Billo:2019fbi}, we exponentiate  (\ref{eq:one-loop in general superconformal theories}) and we interpret the contribution of the one-loop determinants as an interaction potential for the matrix model, i.e. 
   \begin{equation}
\label{eq:Sint}
\left|Z^\mathcal{R}_{\text{1-loop}}\right|^2 = \ee^{-S_{\rm Int}}(a)  \ , \quad \text{where} \quad  S_{\rm Int} 
=\Tr_{\mathcal{R}}^\prime\log H(\mathrm{i}a) \ ,
\end{equation} and we recall that the primed trace is  $\Tr_{\mathcal{R}}^\prime=\left(\Tr_{\cR}-\Tr_{\rm Adj}\right)$. When $\cR=\rm Adj$, the theory reduces to $\mathcal{N}=4$ SYM and the matrix model becomes purely Gaussian\footnote{As showed in \cite{Pestun:2007rz}, in $\mathcal{N}=4$ SYM also the  contribution of the instantons is trivial when $\cR=\rm Adj$.}. 

 For general setups, however, the primed trace is non-vanishing and precisely describes the matter sector of the \textit{difference theory}, which arises
when we subtract the field content of $\cN$ = 4 SYM from that of  $\cN = 2$ theories
with hypermultiplets in the representation $\cR$ \cite{Andree:2010na,Billo:2019fbi}. From the perturbative field theory point of
view, the matrix model suggests to construct the interaction contributions by considering
the diagrams characterized by internal lines in the representation $\cR$ and by subtracting
identical terms in which $\cR = \rm Adj$. This is precisely the pattern with which we organized perturbation theory in flat space (see eq.\1(\ref{eq:pert expansion})).  

%
%
   
	\subsection{The $\frac{1}{4}$-BPS longitude Wilson loop}
	\label{sec:mm1/4}
Let us now turn our attention to the $\frac{1}{4}$-BPS wedge Wilson loop (\ref{eq:1/4 Wilson loop}). In the matrix model approach, we propose the following representation for the operator
   \begin{equation}
		\label{eq:matrix repr wedge}
	W(a,\alpha) = \dfrac{1}{N} \tr \exp(2\pi a \alpha) \  ,
	\end{equation}  where  $\alpha$ is defined in eq. (\ref{eq:vew1/4 N4 and its expansion}) and involves the opening angle $\delta$. Note that the previous expression is a simple modification of the usual phase factor describing the $\frac{1}{2}$-BPS Wilson loop \cite{Pestun:2007rz} and reduces to it in the limit  $\delta\to \pi$.

The vacuum  expectation value of  (\ref{eq:matrix repr wedge}) can be computed  as follows
 \begin{align}
\label{eq:v.e.v part 2}
\left<W\right>
&=\dfrac{1}{\cZ} \int \dd a \ \ee^{-\tr a^2-S_{\rm int}(a,g)} W(a, g\alpha)= \dfrac{\left< \ee^{-S_{\rm int}(a,g)} W_{}(a, \alpha g) \right>_0}{\left< \ee^{-S_{\rm int}(a,g)}  \right>_0}  \ ,
\end{align} where the subscript $0$ denotes the expectation value in the Gaussian matrix model. To obtain the previous expression, we rescaled the integration variable to have a unit coefficient in front of the Gaussian term and normalized the integration measure as $\int \dd a \ee^{-\tr a ^2} = 1 $. The operator and the interaction action take the following form 
\begin{align}
	\label{eq:rescaled op}
W(a,\alpha g)&= \dfrac{1}{N} \tr \exp(a\dfrac{\alpha g }{\sqrt{2}}) = 1 + \dfrac{\alpha^2 g^2 }{4N}\tr a ^2 +\cO(g^4)	\ , \\[0.4em]
	\label{eq:Sint rescaled}
	{S}_{\rm int}(a,g) &=-  \sum_{p=2}^{\infty}\left(-\dfrac{g^2}{8\pi^2}\right)^{p}\frac{\zeta(2p-1)}{p}\Tr_\cR^\prime a^{2p}\  \ .
\end{align}

In $\mathcal{N}=4$ SYM, where the matrix model is purely Gaussian, eq. (\ref{eq:v.e.v part 2}) is consistent with the localization results obtained by Pestun in \cite{Pestun:2009nn, Bassetto:1998sr, Giombi:2009ms, Giombi:2009ds} and it precisely coincides with eq. (\ref{eq:vew1/4 N4 and its expansion}). Indeed, expanding the operator (\ref{eq:rescaled op}), we find that 
\begin{align}
	\label{eq:W0}
	\left<W\right>_{\mathcal{N}=4}
	& = \int\dd{a} \ee^{-\tr a^2}W(a,\alpha g )\notag	\\[0.4em]
	&=1 + \frac{(g\alpha)^2 }{4} \frac{\vvev{\tr a^2}_0}{N} + \frac{(g\alpha)^4}{4! 2^2} \frac{\vvev{\tr a^4}_0}{N} 
	+    \cO((g\alpha)^6)  \ . 
\end{align}
 By evaluating the correlators via the recursive relations of \cite{Billo:2017glv}, we verify that the previous expression coincides with the perturbative expansion of eq.\1(\ref{eq:vew1/4 N4 and its expansion}).

To proceed with the calculation of (\ref{eq:v.e.v part 2}), we expand the operator in powers of $g$, i.e. 
\begin{equation}
	\label{eq:v.e.v part 3}
	\begin{split}
			\left<W\right> &= \frac{1}{N} \sum_{k=0}^{\infty}\dfrac{(g\alpha)^{k}}{k!2^{\frac{k}{2}}} \frac{\left< \ee^{-S_{\rm int}(a,g)} \tr a^{2k} \right>_0}{\left< \ee^{-S_{\rm int}(a,g)}  \right>_0} 
		=	\left<W\right>_{\mathcal{N}=4} +
		 \frac{1}{N} \sum_{k=0}^{\infty}\dfrac{(g\alpha)^{k}}{k!2^{\frac{k}{2}}}\mathsf{\Pi}_k(g)\ ,
	\end{split}
\end{equation} where the quantities $\mathsf{\Pi}_k(g) $ represent quantum corrections resulting from the interaction potential $S_{\rm int}(a,g)$. Expanding the interaction actions in eq.\1(\ref{eq:v.e.v part 3}) via (\ref{eq:Sint rescaled}), we find that 
  \begin{equation}
\label{eq:qauntum corrections matrix model}
	\begin{split}
\mathsf{\Pi}_k(g) 
&= \left(\frac{g^2}{8\pi^2}\right)^2\dfrac{\zeta(3)}{2}\big<\tr a^k\Tr_\cR^\prime a^4\big>_{0,\rm c} -\left(\dfrac{g^2}{8\pi^2}\right)^3 \dfrac{\zeta(5)}{3} \big<\tr a^k\Tr_\cR^\prime a^6\big>_{0,\rm c} + \cO(g^{8}) \ ,
\end{split}
\end{equation}where the subscript $0,\rm c $ denotes the \textit{connected correlator} in the Gaussian matrix model. Subleading corrections in the previous expression involve Riemann zeta functions with higher transcendentality. Finally, we  note that using eq. (\ref{eq:flat integration measure matrix model}), we can express the primed traces in terms of cyclic tensors, i.e.  \begin{equation}
\label{eq:general primed trace}
\Tr_{\cR}^\prime a^{2k} = C^\prime_{b_1\ldots b_k}  a^{b_1}\ldots a^{b_k} \ , \quad \text{where} \quad  C^\prime_{b_1\ldots b_k}=\Tr_{\cR}^\prime T_{b_1}\dots T_{b_k}  \ . 
\end{equation}
 

In order to test the matrix model predictions with the perturbative results obtained in flat space at three loops, we need the explicit expression of $\mathsf{\Pi}_2(g)$, i.e. 
 \begin{equation}
	\begin{split}
		\mathsf{\Pi}_2(g)&=	\left(\frac{g^2}{8\pi^2}\right)^2\dfrac{\zeta(3)}{2}\big<\tr a^2\Tr_\cR^\prime a^4\big>_{0,\rm c} 
		-\left(\dfrac{g^2}{8\pi^2}\right)^3 \dfrac{\zeta(5)}{3} \big<\tr a^2\Tr_\cR^\prime a^6\big>_{0,\rm c} + \cO(g^{8}) \ . \\ 
	\end{split}
\end{equation} For an arbitrary representation $\cR$, we can evaluate the connected correlators in the previous expression by Wick theorem. Introducing the free propagator  $\big<a^a a^b\big>_0 =\delta^{ab}$,  we find that 
\begin{equation}
		\label{eq:tra^2}
			\mathsf{\Pi}_2(g)=	NC_F\left(\dfrac{g^2}{8\pi^2}\right)^2\zeta(3)6\cC^\prime_{4} -NC_F \left(\dfrac{g^2}{8\pi^2}\right)^3 \zeta(5) 30\cC^\prime_6 + \cO(g^8)  \ ,
\end{equation} where we recall that the coefficient $\cC_{4}^\prime$ is defined in (\ref{eq:vev at three-loop def}) and we introduced \begin{equation}
\label{eq:c6}
\Tr_{\cR}^\prime T^a T^b T^d T^d T^e T^e = \delta_{ab}\cC^\prime_6 \ ,
\end{equation}
 where we recall that $\Tr_{\cR}^\prime=\Tr_{\cR}-\Tr_{\rm Adj}$. By combining together eq.s (\ref{eq:tra^2}) and (\ref{eq:v.e.v part 3}), we finally find the three-loop prediction of the matrix model is  
\begin{equation}
\label{eq:three-loop correction matrix model}
	\left<W\right>  =		\left<W\right>_{\mathcal{N}=4}\big|_{g^6}+g^6\alpha^2 \dfrac{C_F 3\zeta(3)\cC_4^\prime}{2^7\pi^2} + \mathcal{O}(g^8) \ . 
\end{equation}
 The previous expression is valid for any superconformal setup and it is in perfect agreement with the flat-space result (\ref{eq:vev at three-loop def}). Let us note that the $\zeta(3)$-correction in the previous expression, resulting from the first connected correlator of the quantum correction $\mathsf{\Pi}_2$ (\ref{eq:tra^2}), is in one-to-one correspondence with the the single-exchange diagrams of eq.\1(\ref{eq:delta6 bubble}).
\subsection{A special class of theories}
Obtaining higher loop corrections with the matrix model approach is quite simple. Conversely, the perturbative calculation in the field theory approach is way more involved.  However, for the so-called $\mathbf{E}$ theories \cite{Billo:2019fbi}, where the matter hypermultiplets  transform in the rank-two symmetric ($S$) and antisymmetric ($A$) representation\footnote{The Dynkin indices of the symmetric and antisymmetric representation are, respectively, given by $i_S=(N+2)/2$ and $i_A=(N-2)/2$. As a result, $i_\cR=i_S+i_A=N$ and consequently, according to eq.\1(\ref{eq:beta-function}), the theory is superconformal. }  of SU($N$), it is actually possible to test  matrix model prediction with ordinary perturbation theory at four loops. The key observation is that in this model, the coefficient $\cC_{4}^\prime$, appearing in eq.\1(\ref{eq:three-loop correction matrix model}), vanishes. This can be proved explicitly by exploiting  the explicit definition of the colour factor\footnote{The explicit expression of the Casimir operators for the symmetric and antisymmetric representation of SU($N$) are given by \begin{equation}
		C_A=\frac{(N+1)(N-2)}{N}\ , \quad 	C_S=\frac{(N-1)(N+2)}{N} \  .
\end{equation} } $\cC_{4}^\prime $ (\ref{eq:vev at three-loop def}). As a result,  the observable coincides with the   $\mathcal{N}=4$ counterpart up to three loops.
Using eq.\1(\ref{eq:qauntum corrections matrix model}), we can easily determine the four-loop corrections. We find that   \begin{equation}
\label{eq:observable in E theory}
\left<W\right>_\mathbf{E} = \left<W\right>_{\cN=4} -g^8\alpha^2\dfrac{C_F\zeta(5)15\cC_6^\prime}{2^{10}\pi^6} \ ,
\end{equation} where we recall that $(g\alpha)=g \alpha$, with $\alpha$ given by eq.\1(\ref{eq:vew1/4 N4 and its expansion}). Note that the effects of the interaction action appear in the observable via a single term proportional to $\zeta(5)$. This correction results from the quantity $\mathsf{\Pi}_2$ (\ref{eq:tra^2}) and we expect, in analogy to the three-loop contribution (\ref{eq:three-loop correction matrix model}), that it arises  from proper single-exchange diagrams in perturbative field theory. We will verify that  eq.\1(\ref{eq:observable in E theory}) actually matches perturbation theory in Appendix\1\ref{sec:four-loop corrections in E theory}.

\section{The case of $\mathcal{N}=2$ SQCD at large-$N$}\label{sec5}

In the previous sections, we proposed a matrix model representation for the expectation value of the wedge operator (\ref{eq:1/4 Wilson loop}). In this section, we will employ this description to determine the large-$N$ limit of the observable in conformal SQCD. 

Let us begin with recalling that this theory  consists of a $\mathcal{N}=2$ vector multiplet coupled to $2N$  hypermultiplets in the fundamental representation of SU($N$). On the four-sphere, we can obtain the partition function  from eq.s\1\eqref{eq:partition function} and \eqref{eq:one-loop in general superconformal theories} setting $\cR= 2N \Box$. 

In the 't Hooft limit, where $N\to\infty$ while $\lambda=g^2 N$ is kept fixed, we find that \cite{Passerini:2011fe}
\begin{equation}
	\label{eq:sqcdmm}
	\cZ=\int \dd^{N-1}a\,\prod_{u<v}(a_u-a_v)^2\,\,\ee^{\frac{-8\pi^2 N}{\lambda}\sum_u a_u^2}\frac{\prod_{u<v}H^2(a_u-a_v)}{\prod_{u}H^{2N}(a_u)} \ , 
\end{equation}
where $a_u$, with $u=1,\ldots N$, denotes the eigenvalues of  the matrix $a$.

The integral in eq.\1(\ref{eq:sqcdmm}) can be analysed by the saddle-point method \cite{Passerini:2011fe}\footnote{See also \cite{Bourgine:2011ie} for an alternative derivation.}. 
Extremising  the effective action \eqref{eq:Sint}, we obtain the saddle-point equation in the  large $N$ limit
\begin{equation}\label{eq:saddle}
	\begin{gathered}
		\frac{8\pi^2}{\lambda}x-K(x)=\mathsf{P}\!\!\int_{-\mu}^{\mu}\dd y\, \rho(y)\left( \frac{1}{x-y}-K(x-y)\right)\, \\  
		x\in(-\mu,\,\mu)\,, \qquad K(x)\equiv-\frac{H'(x)}{H(x)}\,.
	\end{gathered}
\end{equation}
In the previous expression, $\mathsf{P}$ denotes the principal value of the integral, $H(x)$ is a function defined in \eqref{H}, $\rho(x)$ is the spectral density, i.e. the continuous limit of the eigenvalue distribution, while $\mu$ is the endpoint of the distribution. The  dependence of $\mu$ on the 't Hooft coupling $\lambda $ is determined by the unit normalization of $\rho$.

The expectation value  of the wedge operator  in the continuous limit reads as
\begin{equation}\label{eq:v.e.v._sqcd}
	\left<W\right>_{\rm SQCD} =\int_{-\mu}^{+\mu} \dd x\,\rho(x)\ee^{2\pi\alpha x}\,,
\end{equation}
where we recall that $\alpha$ is defined in (\ref{eq:vew1/4 N4 and its expansion}). Therefore, the first task is to the determine the eigenvalue distribution  from eq.\1\eqref{eq:saddle}. 
Unfortunately, an exact expression which is valid for every value of $\lambda$  as in the maximally supersymmetric case \cite{Erickson:2000af, Drukker:2000rr} is not available. Nevertheless, we will derive explicit results at weak and strong coupling.

\subsection{Weak coupling}
To begin with, we compute the wedge in the small 't Hooft coupling limit, i.e.  $\lambda\ll1$. Following \cite{Passerini:2011fe}, we rewrite the saddle-point equation as 
\begin{equation}\label{eq:rho}
	\rho(x)=\frac{8\pi}{\lambda}\sqrt{\mu^2-x^2}-\frac{1}{\pi^2}\mathsf{P}\int_{-\mu}^\mu\frac{\dd y}{x-y}\sqrt{\frac{\mu^2-x^2}{\mu^2-y^2}}\int dz\rho(z)(K(y-z)-K(y))\,.
\end{equation}
The first term on the r.h.s. is the Wigner semi-circle distribution of eigenvalues  in  $\mathcal{N}=4$ SYM \cite{Erickson:2000af, Drukker:2000rr}. As $\lambda\ll1$, we can also take $\mu\ll1$. Hence, we can approximate $K(x)$ with the first orders of its power expansion
\begin{equation}
	K(x)\simeq 2\zeta(3)x^3-2\zeta(5)x^5 \ .
\end{equation} 
We determine $\rho$ by solving \eqref{eq:rho} iteratively. A straightforward calculation reveals that the density takes the following form
\begin{equation}
	\rho(x)=\dfrac{256\pi}{\lambda\mu^4}\left(\frac{
		5 \mu ^4 \zeta (5) \left(2
		x^2-\mu ^2\right)-4}{ \left( 25
		\mu ^6 \zeta (5) \left(\mu ^6 \zeta
		(5)-16\right)+96  \zeta
		(3)-128\right)}\right) \sqrt{\mu ^2-x^2} \ .
\end{equation}
To determine $\mu$ as a function of $\lambda$, it is sufficient to require that $\rho$ is normalized to one.  As a result, one can easily verify that  at weak coupling the correct relation is given by  \begin{equation}
	\mu = \dfrac{\sqrt{\lambda}}{2\pi}-\dfrac{3\lambda^{5/2}\zeta(3)}{256\pi^5}+\dfrac{5\lambda^{7/2}\zeta(5)}{512\pi^7} + \frac{63 \lambda^{9/2} \zeta (3)^2}{65536
		\pi ^9} + \dots \ .
\end{equation} 
Evaluating the matrix model average for $\lambda\ll1$, we find the following result
\begin{equation}\label{eq:W-weak}
	\begin{split}
		\left<W\right>_{\rm SQCD} =&\int_{-\mu}^{\mu}\dd{x}\rho(x)\ee^{2\pi\alpha x} = 
		1+\frac{\alpha ^2 \lambda }{8}+\frac{\alpha
			^4 \lambda ^2}{192}+\lambda ^3
		\left(\frac{\alpha ^6}{9216}-\frac{3
			\alpha ^2 \zeta (3)}{512 \pi
			^4}\right) +\\
		&\lambda ^4 \left(\frac{\alpha
			^8}{737280}-\frac{\alpha ^4 \zeta
		(3)}{2048 \pi ^4}+\frac{15 \alpha ^2
			\zeta (5)}{4096 \pi
			^6}\right)+
		\cO\left(\lambda ^5\right) \ .
	\end{split}
\end{equation}
The previous expression matches the three-loop prediction   eq.\1\eqref{eq:three-loop correction matrix model}, once we specialize the latter  to  conformal SQCD and take the large-$N$ limit.

\subsection{Strong coupling}
In the weak coupling regimes, the effects of the interaction action, encoded in the function $K(x)$, plays the role of a small perturbation of the Wigner distribution and they do not change its behavior. Conversely, in the strong coupling regimes, $K(x)$ drastically modifies the distribution, introducing a repulsive force which drives the endpoint $\mu$ to infinity. 

%
%
%
%

This unusual behaviour makes the computation of Wilson loops subtle in the limit $\lambda\to\infty$. To see this, we begin with observing that  \cite{Passerini:2011fe} 
\begin{equation}\label{eq:distrinf}
	\rho_\infty(x)=\frac{1}{2\cosh\left(\frac{\pi x}{2}\right)}\,.
\end{equation}
To evaluate the Wilson loop, we simply have to average (\ref{eq:matrix repr wedge})  over $\rho_\infty(x)$. If $\alpha<1/4$, the integral is well-defined and yields a  $\lambda$-independent result
\begin{equation}
	\label{eq:alpha<1/4}
	\left<W\right>_{\rm SQCD}=\int_{-\infty}^\infty \dd x\, \ee^{2\pi \alpha x}\rho_{\infty}(x)=\frac{1}{\cos2\pi\alpha }\, .
\end{equation}
Conversely, for $\alpha \geq 1/4$, it is \textit{naively} divergent. This case includes the value $\alpha = 1$ for which the wedge operator coincides with the circular configuration. 
To avoid this infinite, we have to carefully take the limit $\lambda\to \infty$. 

\begin{figure}[h]
	\centering
	\includegraphics[width=0.5\textwidth]{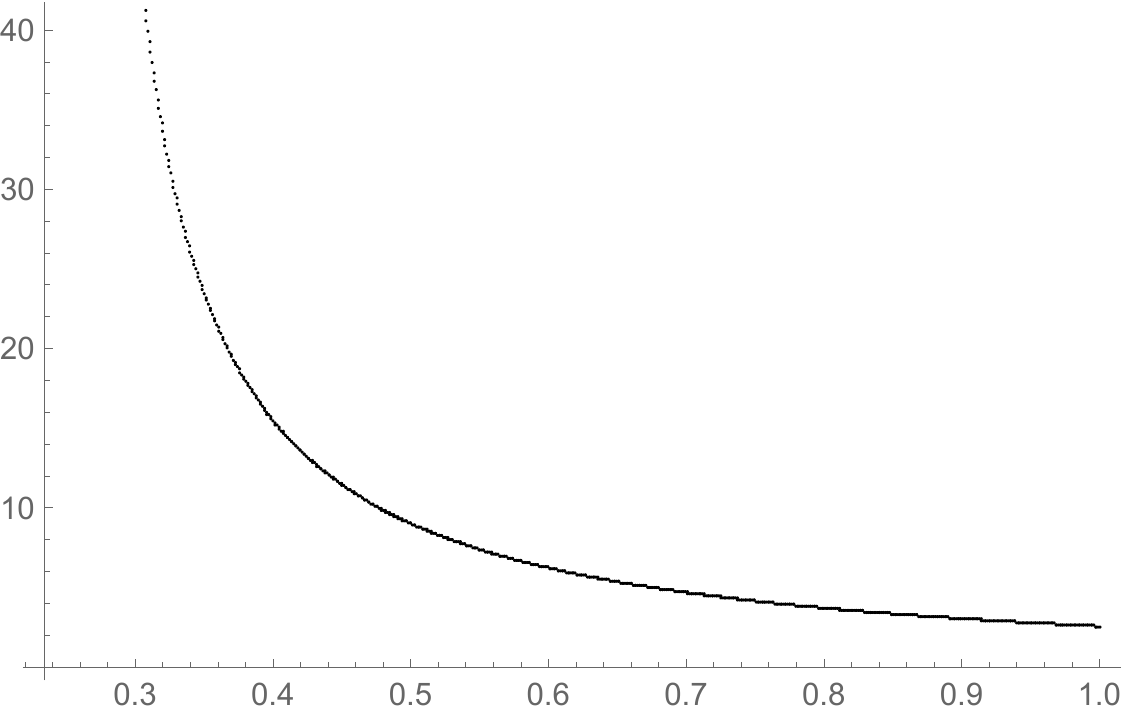}
	\caption{We numerically evaluate $R_\alpha$ for the discrete values between $1/4\le\alpha\le1$.}
	\label{fig1}
\end{figure}

To do so, we assume  $\lambda$ to be large but not infinite. As a result, the Fourier transform of the spectral density takes the following form \cite{Passerini:2011fe} 
\begin{equation}
	\hat{\rho}(\omega)=\frac{1}{\cosh\omega}+\frac{2\sinh^2\frac{\omega}{2}}{\cosh\omega}F(\omega)-G_-(\omega)\ee^{\mathrm{i}\mu\omega}\sum_{n=1}^{+\infty}\frac{n\, r_ne^{-\mu\nu_n}F(-\mathrm{i}\nu_n)(\omega-\mathrm{i}\pi/2)}{(n+1)(\omega+\mathrm{i}\nu_n)(\omega+\mathrm{i}\pi/2)}\,, 
\end{equation} 
where
\begin{equation}
	\begin{gathered}
		G_-(\omega)=\frac{\sqrt{8\pi^3}2^{-\mathrm{i}\omega/\pi} \Gamma\left(\frac{1}{2}+\frac{\mathrm{i}\omega}{\pi}\right)  }{\omega\,\Gamma\left(\frac{\mathrm{i}\omega}{2\pi}\right)^2}\,, \quad
		\nu_n=\pi (n+1/2)\,,\\
		r_n=\frac{(-2)^{n+1}\Gamma\left(\frac{n}{2}+\frac{5}{4}\right)^2}{\sqrt\pi (n+1/2)\Gamma(n+1)}\, ,
	\end{gathered}
\end{equation}
while $F(\omega)$ is the Fourier transform of the second term on the r.h.s. of eq.\1\eqref{eq:rho}.
For the moment, we consider only its leading term, well-approximated by a Bessel function. We only need the explicit expression for imaginary arguments and large $\mu$ value, that is
\begin{equation}
	F(-\mathrm{i}x)=\frac{8\pi^2\mu I_1(\mu x)}{\lambda x}\simeq \frac{\ee^{\mu x}}{\lambda}\sqrt{\frac{32\pi^3 \mu}{\nu_n^3}}\,.
\end{equation}
We can now compute the v.e.v. of the Wilson loop by analytically continuing $\hat{\rho}(\omega)$ for imaginary values of $\omega$.
Specifically, evaluating $\hat\rho(-2\pi \mathrm{i}\alpha)$,  we can express the result as 
\begin{equation}\label{eq:Walpha}
\left<W\right>_{\rm SQCD}= R_{\alpha} \frac{ \sqrt{\mu }}{\lambda }\exp \left(2 \pi  \alpha  \mu \right) \ ,
\end{equation}
with $R_\alpha$ being  given by
\begin{equation}
	\label{eq:Ra def}
	\begin{gathered}
		R_\alpha=\frac{\alpha2^{2(2-\alpha)}\,\Gamma\left(\frac{1}{2}+2\alpha\right)(4\alpha+1)}{\sqrt{\pi}\Gamma(1+\alpha)^2(4\alpha-1)}
		\sum_{n=1}^{+\infty}\frac{n\,r_n}{(n+1)\left(n+\frac{1}{2}\right)^{3/2}(4\alpha-2n-1)}
		-\frac{4}{\alpha^{\frac{3}{2}}}\frac{\sin^2\pi\alpha}{\cos 2\pi\alpha}\ .
	\end{gathered}
\end{equation}
We observe that for $\alpha=1$ the previous expression coincides with the result derived in \cite{Passerini:2011fe} and, for $\alpha>1/4$, is finite since the two apparent poles at $\alpha=3/4$ cancel each other out. Moreover, at the critical value $\alpha=1/4$, $R_\alpha$ diverges (see Fig \ref{fig1}).

Let us note that one can express \eqref{eq:Walpha} as a function of only $\lambda$. To do so, we impose the unit normalization of $\rho$ as in \cite{Passerini:2011fe}
\begin{equation}\label{eq:Cdef}
	C\sqrt{\mu}\,\ee^{\frac{\pi}{2}\mu}=\lambda\,,\qquad \mathrm{with} \qquad C=\frac{16}{r_0}\sum_{n=0}^{+\infty}\frac{r_n}{(2n+1)^{3/2}(n+1)}\, .
\end{equation}
Finally, replacing $\mu$ with $\lambda$ in \eqref{eq:Walpha}, we obtain
\begin{equation}\label{eq:Wlambda}
	\left<W\right>_{\rm SQCD} =\frac{R_\alpha}{C}\left(\frac{\pi}{2\log\left(\frac{\lambda}{C}\right)}\right)^{2(\alpha-1/4)}\left(\frac{\lambda}{C}\right)^{4(\alpha-1/4)}\, .
\end{equation}
For $\alpha>1/4$, the previous expression exhibits a the power law dependence analogous to that originally observed in \cite{Rey:2010ry, Passerini:2011fe}.

At this stage, it is interesting to compare the last expressions with eq.\1 \eqref{eq:alpha<1/4}. We observe that the observable exhibits a singular behavior at the critical value $\alpha=1/4$ or, equivalently, at the opening angle 
\begin{equation}
	\label{eq:critical angle}
	\delta=\frac{1}{4} \left(4-\sqrt{15}\right) \pi\,.
\end{equation}
This behavior in the parameter $\alpha$ is a new phenomenon for the wedge Wilson loop.

\paragraph{Subleading corrections}

\begin{figure}
	\centering
	\includegraphics[width=0.5\textwidth]{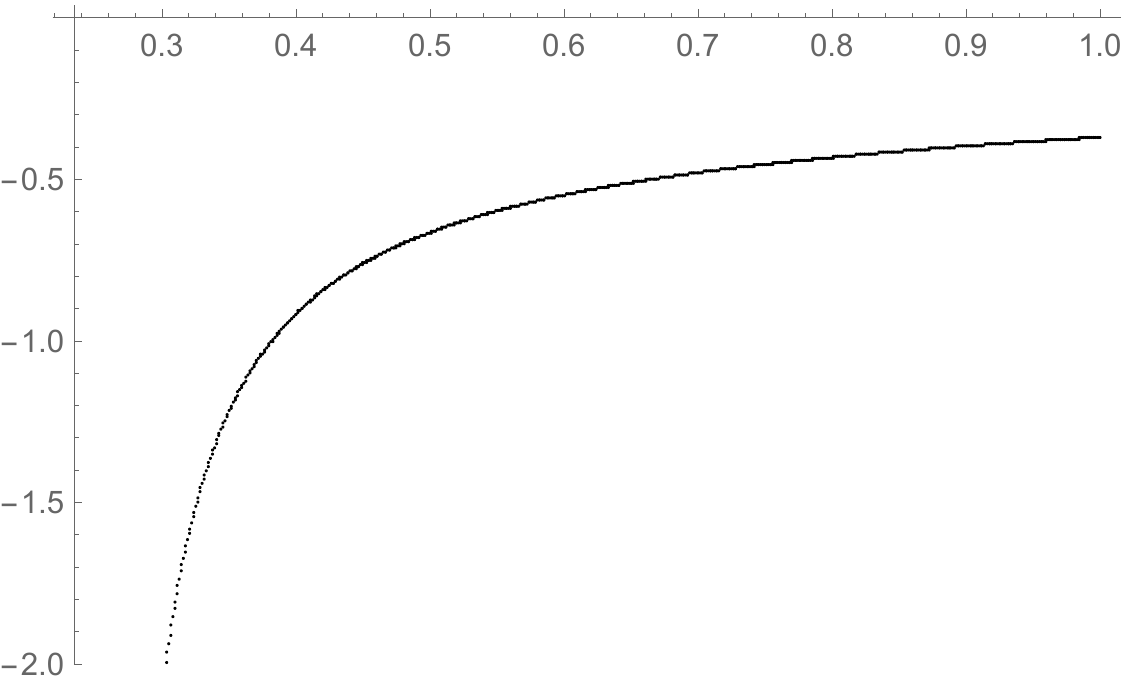}
		\caption{We numerically plot $\delta R_\alpha$ as we did for $R_\alpha$. Notice that for $\alpha$ away from the singularity, the numerical value are lower by a factor 10.}
		\label{fig5.2}
\end{figure}

Eq. (\ref{eq:Ra def}) captures the leading order of the observable at large 't Hooft coupling. Subleading contributions can be calculated by using the results outlined in Appendix D of \cite{Passerini:2011fe}. For instance, if we include the first subleading correction to the function $F$ for purely imaginary values, i.e. 
\begin{equation}  
	\delta F(-\mathrm{i}x)=-a\sqrt{\frac{2}{\pi x}}\frac{\sqrt{\mu}}{\lambda}\ee^{\mu x}\,, 
\end{equation}
where \begin{equation}
	a=2^{\frac{7}{2}}\pi^{-\frac{3}{2}} \sum_{m,n=1}^{+\infty} \frac{mr_m}{(2m+1)^{\frac{3}{2}}(m+1)}\dv{n}\left[ \frac{\Gamma\left(2n+\frac{3}{2}\right)}{2^{2n} n^{\frac{3}{2}}\Gamma^2(n)\left(2n-\frac{1}{2}\right)\left(2n-m-\frac{1}{2}\right) }\right]\,, 
\end{equation}
we find that the corresponding  correction to the Wilson loop takes the form 
\begin{equation}
	\delta R_\alpha=\frac{2 a\sin ^2(\pi  \alpha)}{\left(\pi  \sqrt{\alpha}\right) \cos (2 \pi  \alpha)}-a\frac{\alpha2^{2(1-\alpha)}\,\Gamma\left(\frac{1}{2}+2\alpha\right)(4\alpha+1)}{\pi^{\frac{3}{2}}\Gamma(1+\alpha)^2(4\alpha-1)}
	\sum_{n=1}^{+\infty}\frac{n\,r_n}{(n+1)\left(n+\frac{1}{2}\right)^{\frac{1}{2}}(4\alpha-2n-1)} \ .
\end{equation}
Note that $\delta R_\alpha$ is singular for $\alpha=1/4$.
From the plot in Fig.\1\ref{fig5.2}, we observe $\delta R_\alpha$ competes with $R_\alpha$, but its absolute value is smaller by an order of magnitude. Thus, the correction $\delta R_\alpha$ does not affect the qualitative behaviour of $R_\alpha$, justifying \emph{a posteriori} having neglected it.

\section{Discussions and Outlook}\label{sec7}
The matrix model on $\mathbb{S}^4$ \cite{Pestun:2007rz} represents one of the most powerful tools to study protected observables in $\mathcal{N}=2$ superconformal gauge theories and in different regimes, such as the large-$N$ limit at strong coupling which plays a crucial  role in the AdS/CFT correspondence.

In this work, we expanded the set of observables which can be studied within this framework, by constructing a family of $\frac14$-BPS Wilson in $\mathcal{N}=2$ superconformal theories, with contours consisting of two meridians  on a two-dimensional sphere $\mathbb{S}^2$. We analysed these observables at the perturbative level, carefully computing their expectation value at three loops by usual  Feynman diagram techniques. Inspired by this calculation, we proposed a simple modification of the matrix model capturing the $\frac{1}{2}$-BPS Wilson loop and showed that it reproduces the perturbative calculation. Our proposal is also consistent, under reasonable assumptions, with previous computations in the specific case of conformal SQCD, as we will discuss shortly.

 In the specific case of conformal SQCD, we studied the observable in the large-$N$ limit at strong coupling, where the theory is expected to be described by a gravitational dual. Hence, our result provides a possible test for a putative dual string, which should reproduce the dependence on the wedge opening angle. Our analysis reveals a surprising  transition in the observable at a certain  critical size of the wedge. For configurations sufficiently close to the circular one, the effective string tension is logarithmic in the 't Hooft coupling and vanishes at the critical angle associated with the transition. Moreover, for smaller values of the opening angle than the critical one, the wedge  becomes independent of the coupling. This is a new phenomenon for the observable since, in the only previously known example, namely $\mathcal{N}=4$ SYM, the observable  behaves as $\ee^{\alpha\sqrt{\lambda}}$ for every value of $\alpha$. 

We conclude this work by discussing the relations of our computations with other results appeared in the literature and presenting possible extensions of this work.

\paragraph{The cusp and the wedge} It is well known that when a loop operator $W$ makes a sudden turn by an angle $\delta$ its v.e.v. develops a logarithmic divergence, known as cusp anomalous dimension \cite{Polyakov:1980ca}. In this case, the observable takes the following form 
\begin{equation}
	\langle W\rangle\simeq\ee^{-\Gamma_{\textup{cusp}}(\delta,\,g)\log\frac{L}{\epsilon}} \ ,
\end{equation}
where $L$ and $\epsilon$ are the IR and the UV cutoff, respectively, while  $\Gamma_{\textup{cusp}}$ stores the relevant physical information.
This discussion can be easily extended to supersymmetric line operators. In these cases, $\Gamma_{\textup{cusp}}$ also depends on an additional internal angle $\varphi$ that weights the different couplings with the scalar fields at the cusp. In general,  the corresponding cusp anomalous dimension will depend on both the physical and internal angles, as well as on the parameters of the theory. However, at the BPS point there are no divergences and consequently, $\Gamma_{\textup{cusp}}=0$ when $\varphi=\pm\delta$.  As a result, it is natural to perform a near-BPS expansion around $\varphi=\delta$, i.e. 
\begin{equation}
	\Gamma_{\textup{cusp}}\simeq (\varphi-\delta)H(\varphi) \ , 
\end{equation}
where $H(\varphi)$ is a function of $\varphi\equiv\delta$ and of theory parameters. 

In  $\mathcal{N}=4$ SYM, $H(\varphi)$ is fully determined by the expectation value of the wedge operator (\ref{eq:1/4 Wilson loop}) according to the formula\footnote{We recall that the opening angle of our cusped Wilson loop (\ref{eq:cusp configuration}) is $\pi-\delta$, see Fig \ref{fig:cusp}. } \cite{Correa:2012at}
\begin{equation}\label{eq:nearbps}
	H(\varphi)=\frac{1}{2}\pdv{\delta}
\left<W\right>_{\mathcal{N}=4}\bigg|_{\delta=\pi+\varphi}\,.
\end{equation}
A natural expectation is that  the same formula applies to $\mathcal{N}=2$ SYM theories, although it is not completely obvious whether it is possible to go through the same analysis of \cite{Correa:2012at}\footnote{A potential technical issue is that some of the manipulations performed in \cite{Correa:2012at} do not preserve enough supersymmetry in the $\mathcal{N}=2$ case.}. 

However, it is interesting to note that using the results of this work,  we can provide a non-trivial test of (\ref{eq:nearbps})  beyond the maximally supersymmetric case. In fact, the cusp anomaly was computed in perturbation theory up three loops for arbitrary angles in $\mathcal{N}=2$ conformal SQCD \cite{Gomez:2018usu}
\begin{equation}
	\Gamma^{\mathcal{N}=2}_{\textup{cusp}}-\Gamma^{\mathcal{N}=4}_{\textup{cusp}}=g^6\frac{3\zeta(3)}{512\pi^6}\frac{N^4-1}{N}\frac{1+\ee^{2\mathrm{i} \delta}-2\ee^{\mathrm{i} \delta}\cos \varphi}{1-\ee^{2\mathrm{i} \delta}}\mathrm{i} \delta \ .
\end{equation} 
Thus, performing the near-BPS expansion we read the value of $H$
\begin{equation}
	H_{\mathcal{N}=2}-H_{\mathcal{N}=4}=g^6\frac{3\zeta(3)}{512\pi^6}\frac{N^4-1}{N} \varphi \ .
\end{equation} 
It is straightforward to reproduce this result by plugging eq.\1\eqref{eq:three-loop correction matrix model}, specialized to  conformal SQCD, into the r.h.s. of eq\1 \eqref{eq:nearbps}. 

\paragraph{The Bremsstrahlung, the wedge, and the squashed sphere}

A particularly relevant limit of the cusp anomalous dimension is the small-angle regime, where $\Gamma_{\textup{cusp}}$ is captured by the Bremsstrahlung function $B$\begin{equation}
	\Gamma_{\textup{cusp}}\simeq-B(g)\delta^2 \ .
\end{equation}
This function has several applications in amplitudes and defect conformal field theory \cite{Correa:2012at}. Moreover, unlike the cusp anomaly, it is sometimes accessible in supersymmetric theories via localization. Thus, it is natural to examine  a direct connection between this quantity and our results for the wedge\footnote{
In $\mathcal{N}=4$ SYM, where the wedge depends only on the rescaled coupling $(g\alpha)=\alpha g$,  it is possible to rewrite eq.\1\eqref{eq:nearbps} as a derivative w.r.t. to the coupling, obtaining  \cite{Correa:2012at}
\begin{equation}\label{eq:b4}
	H^{\mathcal{N}=4}=\frac{2\delta}{1-\delta^2/\pi^2} B(g\alpha) \ .
\end{equation} 
The Bremsstrahlung function therefore determines the near-BPS limit of the cusp. 
We remark that this result is strictly valid in $\mathcal{N}=4$ SYM since the operator realizing the marginal deformations, related to the derivatives w.r.t. to the coupling constant $g$, is part of the stress tensor multiplet. The latter property does not apply to $\mathcal{N}=2$ theories, preventing us from directly extending eq.\1 \eqref{eq:b4} to the less supersymmetric case.}. Specifically, we observe that assuming the formula
\begin{equation}\label{eq:bwedge}
	B=-\frac{1}{4} \left.\frac{\partial^2}{\partial\delta^2}\log\langle W\rangle\right|_{\delta=\pi}\,,
\end{equation}
we recover the three-loop Bremsstrahlung, derived in \cite{Gomez:2018usu}. Moreover, at strong coupling, using eq.\1\eqref{eq:Walpha}, we get the result of \cite{Fiol:2015mrp}.

We can connect our proposal (\ref{eq:bwedge}) to another exact expression available in the literature for the Bremsstrahlung, which is valid for any $\mathcal{N}=2$ theory. This formula, conjectured in \cite{Fiol:2015spa} and proved in \cite{Bianchi:2019dlw}, relates $B$ to the circular Wilson loop computed on the squashed sphere $S^4_b$, namely a deformation of the sphere depending on a real parameter $b$.
For $b=1$,  $S^4_b$ reduces to $\mathbb{S}^4$. In particular, the Bremsstrahlung is proportional to the derivative of the circular Wilson loop w.r.t. to $b$
\begin{equation}
	\label{eq:B}
	B=\frac{1}{4\pi^2}\pdv{b}\log\langle W_{\frac12}\rangle_b \bigg|_{b=1} \ .
\end{equation}
This formula, combined with localization, provides a concrete way to compute $B$ in Lagrangian theories. In fact,  the  circular Wilson loop v.e.v. on the squashed sphere admits the following matrix model representation \cite{Hama:2012bg}
\begin{equation}
	\label{eq:wls4b}
	\langle W_{\frac12}\rangle_b=\int \mathcal{D}a \,\ee^{-\tr a^2} |Z_{\textup{1-loop}}(a,\,b) Z_{\textup{inst}}(a,\,b)|^2 \tr(\ee^{b\frac{g}{\sqrt{2}}a}) \ ,
\end{equation}
where all the different factors reduce to those of the four-sphere as $b\to 1$. According to \cite{Bianchi:2019dlw}, the terms resulting from the derivative of the matrix model do not affect $B$ except for the factor $\tr \ee^{b\frac{g}{\sqrt{2}}a}$ associated with the Wilson loop. Thus,  identifying $b$ with $\alpha$, defined in eq.\1(\ref{eq:vew1/4 N4 and its expansion}), the matrix model (\ref{eq:wls4b}) coincides with our proposal for the expectation value of the wedge (\ref{eq:v.e.v part 3}). Under this identification,  eq.\1(\ref{eq:bwedge}) ties nicely in with eq.\1(\ref{eq:B}), recovering explicitly the $g^8\zeta(5)$ term obtained in \cite{Bianchi:2019dlw}.

\paragraph{Strong coupling and string theory}

Conceivably, the discontinuity we found at strong coupling is an artefact of the large-$N$ limit, as it happens for the Gross-Ooguri phase transition in correlators of Wilson loops \cite{Gross:1998gk, Zarembo:1999bu}.
Even if the origin of this behaviour remains unclear at this stage of the work,  a possible interpretation results from the analysis of high rank representation $\frac12$-BPS Wilson loop in conformal  SQCD  \cite{Fraser:2011qa}. For the $k$-antisymmetric representation, a similar transition occurs between a $\lambda$-independent and $\lambda$-dependent expression of the observable but, in this case, the controlling  parameter is the ratio $k/N$. In this work, the authors speculated on a possible interpretation in the AdS/CFT framework.
In this paradigm, where $\mathcal{N}=4$ SYM is dual to a type IIB string theory on the AdS$_5\times \mathbb{S}^5$ background, Wilson loops in the $k$-antisymmetric representation are dual to D5 branes on AdS$_2\times \mathbb{S}^4$ \cite{Gomis:2006sb}, with the D5's worldvolume acting as a probe of  the internal geometry of the $\mathbb{S}^5$\footnote{For instance, the $k$-symmetric Wilson loops are dual to D3 branes on AdS$_2\times \mathbb{S}^2\subset$ AdS$_5$ and do not experience such a crossover.}. It is therefore tempting to assume that, in the holographic description of  conformal SQCD, the D5 branes probe a compact part of the dual setup, which remains non-geometric as described in \cite{Gadde:2009dj}. The analysis of the wedge corroborates this interpretation. In fact, the dual description of the wedge in $\mathcal{N}=4$ SYM  involves a worldsheet extending in the internal directions, as the D5. Then, a putative holographic dual of conformal SQCD should also explore this region, which remains highly curved. 

\paragraph{Outlook}
There are of course various extensions of this work. First, it would be interesting to further test our proposal at higher orders in perturbation theory and, more ambitiously,  develop a  \emph{bona fide} localization procedure for this class of operators, in analogy to that presented in \cite{Pestun:2009nn} for $\frac18$-BPS Wilson loops in $\mathcal{N}=4$ SYM. Furthermore,  the deep connection between the wedge operator and the Bremsstrahlung function in $\mathcal{N}=4$ SYM, makes natural to further investigate a similar relation also in $\mathcal{N}=2$ theories.  Secondly, it would be interesting to explore the strong coupling limit for high rank representation  wedges, both in conformal SQCD and in the interpolating quiver theory \cite{Zarembo:2020tpf}. Another interesting direction  consists of making contact with the matrix model techniques developed in \cite{Beccaria:2023kbl,Beccaria:2023qnu} for  computing non-planar corrections in special superconformal $\mathcal{N}=2$ SYM theories. Enlarging the class of observables that can be studied by these approaches is a new a source of information for this broad program.

\acknowledgments{
We thank M. Billo', F. Galvagno , G. P. Korchemsky and D. Seminara for inspiring discussions. A.T. is grateful to the Institut de Physique Théorique (CEA) for the kind hospitality during part of this work.
 This work has been supported in part by the Italian Ministero dell’Universit\`a e Ricerca (MIUR), and by Istituto Nazionale di Fisica Nucleare (INFN) through the ``Gauge and String Theory'' (GAST) research project. The work of L.G. has been supported by the OPUS grant no. 2022/47/B/ST2/03313 “Quantum geometry and BPS states” funded by the National Science Centre, Poland.} 

	\appendix

\section{Conventions and algebras}
\label{app2}
This appendix  outlines the spinor conventions and describes the four dimensional $\mathcal{N}=2$ algebra $\mathfrak{su}(2,2|2)$ preserved by the $\frac{1}{4}$-BPS loop operators (\ref{eq:cusp configuration}). 
\subsection{Spinor conventions}
Our conventions for spinors  follows those of \cite{Hama:2012bg}. In Euclidean space the spin group is $\mathrm{Spin}(4)\simeq \mathrm{SU}(2)_{\alpha}\otimes \mathrm{SU}(2)_{\dot{\alpha}}$.  Chiral spinors carry  undotted indices  $\alpha, \beta,\ldots$, while  anti-chiral spinors are characterized by  dotted indices $\dot{\alpha},\dot{\beta},\dots\ $. Note that in Euclidean space, chiral and anti-chiral spinors are not related by complex conjugation.

Spinors are contracted as follows 
\begin{equation}
	\psi\chi\equiv\psi^{\alpha}\chi_{\alpha} \ , \quad \quad \bar{\psi}\bar{\chi}\equiv\bar{\psi}_{\dot{\alpha}}\bar{\chi}^{\dot{\alpha}} \ .
\end{equation} In the following, we raise and lower indices as follows 
\begin{equation}
	\psi^{\alpha}=\epsilon^{\alpha \beta}\psi_{\beta}, \quad \quad \bar{\psi}^{\dot{\alpha}}=\epsilon^{\dot{\alpha}\dot{\beta}}\bar{\psi}_{\dot{\beta}} \ ,
\end{equation}
where $\epsilon^{12}=\epsilon_{21}=\epsilon^{\dot{1}\dot{2}}=\epsilon_{\dot{2}\dot{1}}=1$. Let us note in Euclidean spacetime spinors satisfy \textit{pseudoreality} conditions, i.e. \begin{equation}
	(\psi_\alpha)^\dagger = \psi^\alpha \ .
\end{equation}

The matrices $(\bar{\sigma}^{\mu})^{\dot{\alpha}\alpha}$ and $(\sigma^{\mu})_{\alpha \dot{\beta}}$ are defined as follows
\begin{equation}
	\label{eq:sigma matrices}
	\sigma^{\mu}=(-\mathrm{i}\vec{\tau},\mathbb{I}) \ , \quad \quad \bar{\sigma}^{\mu}= (\mathrm{i}\vec{\tau}, \mathbb{I}) \ ,
\end{equation}
where $\vec{\tau}$ are the ordinary Pauli matrices. Furthermore, these matrices are such that
\begin{equation}(\bar{\sigma}^{\mu})^{\dot{\alpha}\alpha}=\epsilon^{\dot{\alpha}\dot{\beta}}\epsilon^{\alpha \beta}(\sigma^{\mu})_{\beta \dot{\beta}}
\end{equation}
and satisfy the Clifford algebra 
\begin{align}
	\label{eq:clifford}
	\sigma^{\mu}\bar{\sigma}^{\nu}+ \sigma^{\nu}\bar{\sigma}^{\mu}&=2\delta^{\mu \nu}\mathbb{I} \ , \\[0.5em] 
	\bar{\sigma}^{\mu}\sigma^\nu + \bar{\sigma}^\nu\sigma^\mu &= 2\delta^{\mu\nu}\mathbb{I} \ .
\end{align} The previous expressions  imply that \begin{equation}
	\label{eq:double trace}
	\Tr \sigma^\mu\bar{\sigma}^\nu=2\delta^{\mu \nu} \ .
\end{equation} 

\subsection{The $\mathcal{N}=2$ superconformal algebra}
The bosonic part of the algebra contains the conformal algebra, $\mathfrak{so}(5,1)$, and the R-symmetry part, $\mathfrak{su}(2)_R\oplus\mathfrak{u}(1)_r$\footnote{We follow the conventions of \cite{Beem:2013sza}. Notice that we use a different notation for the R-symmetry generators, namely $(R_{\textup{there}})^I_J=(R_{\textup{here}})^I_J+\delta^I_J\frac{r}{2}$. That is, we work with $\mathfrak{su}(2)_R$ and $\mathfrak{u}(1)_r$ separately rather than $\mathfrak{u}(2)_R$.}.
The spacetime generators are $P_{\alpha\dot\alpha}\equiv P_\mu\sigma^\mu_{\alpha\dot\alpha}$, $K^{\dot\alpha\alpha}\equiv K^\mu\bar\sigma_\mu^{\dot\alpha\alpha}$, $D$, ${M_\alpha}^\beta\equiv\frac{1}{4}{(\sigma_\mu\bar{\sigma}_\nu)_\alpha}^\beta M^{\mu\nu}$, ${M^{\dot\alpha}}_{\dot\beta}\equiv -\frac{1}{4}{(\bar{\sigma}_\mu\sigma_\nu)^{\dot\alpha}}_{\dot\beta} M^{\mu\nu}$, generating translations, special conformal transformations, dilations, self and anti-self dual part of the Lorentz group, respectively.
The other factors are generated by ${R^I}_J$ for $\mathfrak{su}(2)_R$ and $r$ for the $\mathfrak{u}(1)_r$ factor.

Their algebra reads as 
\begin{align}
	\comm*{{M_\alpha}^\beta}{{M_\gamma}^\delta}&=\delta_\gamma^\beta{M_\alpha}^\delta-\delta_\alpha^\delta{M_\gamma}^\beta\,, 
	&\comm*{{M^{\dot\alpha}}_{\dot\beta}}{{M^{\dot\gamma}}_{\dot\delta}}&=\delta^{\dot\alpha}_{\dot\delta}{M^{\dot\gamma}}_{\dot\beta}-\delta^{\dot\gamma}_{\dot\beta}{M^{\dot\alpha}}_{\dot\delta}\,,\\
	\comm*{{M_\alpha}^\beta}{P_{\gamma\dot\gamma}}&=\delta_\gamma^\beta P_{\alpha\dot\gamma}-\frac{1}{2}\delta_{\alpha}^\beta P_{\gamma\dot\gamma}\,, 
	&\comm*{{M^{\dot\alpha}}_{\dot\beta}}{P_{\gamma\dot\gamma}}&=\delta^{\dot\alpha}_{\dot\gamma}P_{\gamma\dot\beta}-\frac{1}{2}\delta^{\dot\alpha}_{\dot\beta}P_{\gamma\dot\gamma}\,,\\
	\comm*{{M_\alpha}^\beta}{K^{\dot\gamma \gamma}}&=-\delta^{\gamma}_{\alpha}K^{\dot\gamma\beta}+\frac{1}{2}\delta^{\beta}_{\alpha}K^{\dot\gamma\gamma}\,, 
	&\comm*{{M^{\dot\alpha}}_{\dot\beta}}{K^{\dot\gamma \gamma}}&=-\delta^{\dot\gamma}_{\dot\beta}K^{\dot\alpha\gamma}+\frac{1}{2}\delta^{\dot\alpha}_{\dot\beta}K^{\dot\gamma\gamma}\,,\\
	\comm*{D}{K^{\dot\gamma\gamma}}&=-K^{\dot\gamma\gamma}\,, 
	&\comm*{D}{P_{\gamma\dot\gamma}}&=P_{\gamma\dot\gamma}\,,\\
	\comm*{K^{\alpha\dot\alpha}}{P_{\beta\dot\beta}}&=\delta^\alpha_\beta\delta^{\dot\alpha}_{\dot\beta}D+\delta_{\beta}^\alpha{M^{\dot\alpha}}_{\dot\beta}+\delta^{\dot\alpha}_{\dot\beta}{M_\beta}^\alpha\,.
\end{align}
For the R-symmetry part, we have
\begin{equation}
	\comm*{{R^I}_J}{{R^K}_L}= \delta_J^K{R^I}_L-\delta_L^I{R^K}_J\,.
\end{equation}

The fermionic part is generated by sixteen symmetries, namely the Poincar\'e supercharges  $Q^I_\alpha$, $\bar{Q}_{I\dot\alpha}$, and the superconformal ones $S_I^\alpha$, and $\bar{S}^{I\dot\alpha}$. Their anticommutators are
\begin{align}
	\acomm*{Q^I_\alpha}{\bar{Q}_{I\dot\alpha}}&=\delta^I_JP_{\alpha\dot\alpha}\,, &\acomm*{S_I^\alpha}{\bar{S}^{J\dot\alpha}}&=\delta_I^JK^{\dot\alpha\alpha},\\
	\acomm*{Q^I_\alpha}{S_J^\beta}&=\frac{1}{2}\delta_\alpha^\beta\delta^I_J(D-r)+\delta^I_J{M_\alpha}^\beta-\delta_\alpha^\beta {R^I}_J\,,\\
	\acomm*{\bar{S}^{J\dot\beta}}{\bar{Q}_{I\dot\alpha}}&=\frac{1}{2}\delta^{\dot\beta}_{\dot\alpha}\delta_I^J(D+r)+\delta_I^J{M^{\dot\alpha}}_{\dot\beta}+\delta_{\dot\alpha}^{\dot\beta} {R^J}_I\,.
\end{align}
Finally, the even-odd part is given by
\begin{align}
	\comm*{{M_\alpha}^\beta}{Q_\gamma^I}&=\delta_\gamma^\beta Q_\alpha^I-\frac{1}{2}\delta_\alpha^\beta Q_\gamma^I\,, 
	&\comm*{{M^{\dot\alpha}}_{\dot\beta}}{\bar{Q}_{I\dot\gamma}}&=\delta_{\dot\beta}^{\dot\gamma}\bar{Q}_{I\dot\alpha}-\frac{1}{2}\delta_{\dot\beta}^{\dot\alpha}\bar{Q}_{I\dot\gamma} \,,\\
	\comm*{{M_\alpha}^\beta}{S_J^\gamma}&=\frac{1}{2}\delta_\alpha^\beta S_J^\gamma-\delta_\alpha^\gamma S_J^\beta\,, 
	&\comm*{{M^{\dot\alpha}}_{\dot\beta}}{\bar{S}^{J\dot\gamma}}&=\frac{1}{2}\delta_{\dot\beta}^{\dot\alpha}\bar{S}^{J\dot\gamma}-\delta_{\dot\beta}^{\dot\gamma} \bar{S}^{J\dot\alpha}\,,\\
	\comm*{D}{Q_\alpha^I}&=\frac{1}{2}Q_\alpha^I\,, 
	&\comm*{D}{\bar{Q}_{I\dot\alpha}}&=\frac{1}{2}\bar{Q}_{I\dot\alpha}\,,\\
	\comm*{D}{S_J^\beta}&=-\frac{1}{2}S_J^\beta\,, 
	&\comm*{D}{\bar{S}^{J\dot\beta}}&=-\frac{1}{2}\bar{S}^{J\dot\beta}\,,\\
	\comm*{K^{\dot\alpha\alpha}}{Q_\gamma^I}&=\delta_\gamma^\alpha\bar{S}^{I\dot\alpha}\,, 
	&\comm*{K^{\dot\alpha\alpha}}{\bar{Q}_{I\dot\gamma}}&=\delta^{\dot\alpha}_{\dot\gamma}S_I^\alpha \,,\\
	\comm*{P_{\alpha\dot\alpha}}{S_J^\gamma}&=-\delta_\alpha^\gamma\bar{Q}_{J\dot\alpha}\,, 
	&\comm*{P_{\alpha\dot\alpha}}{\bar{S}^{J\dot\gamma}}&=-\delta_{\dot\alpha}^{\dot\gamma}Q^J_\alpha\,,\\
	\comm*{r}{Q_\alpha^I}&=-\frac{1}{2}Q_\alpha^I\,, 
	&\comm*{r}{\bar{Q}_{I\dot\alpha}}&=\frac{1}{2}\bar{Q}_{I\dot\alpha}\,,\\
	\comm*{r}{S_J^\beta}&=-\frac{1}{2}S_J^\beta\,, 
	&\comm*{r}{\bar{S}^{J\dot\beta}}&=\frac{1}{2}\bar{S}^{J\dot\beta}\,,\\
	\comm*{{R^I}_J}{Q_\alpha^K}&=\delta^K_JQ_\alpha^I-\frac{1}{2}\delta^I_JQ_\alpha^K \,, 
	&\comm*{{R^I}_J}{\bar{Q}_{K\dot\alpha}}&=\frac{1}{2}\delta^I_J\bar{Q}_{K\dot\alpha}-\delta^I_K\bar{Q}_{J\dot\alpha} \,,\\
	\comm*{{R^I}_J}{S_K^\beta}&=\frac{1}{2}\delta^I_JS_K^\beta-\delta^I_KS_J^\beta\,, 
	&\comm*{{R^I}_J}{\bar{S}^{K\dot\beta}}&=\delta^K_J\bar{S}^{I\dot\beta}-\frac{1}{2}\delta^I_J\bar{S}^{K\dot\beta}\,.
\end{align}

\section{Four-loop corrections in the $\mathbf{E}$ theory}
 \label{sec:four-loop corrections in E theory}
In section\1\ref{sec:the matrix model}, we showed that the matrix model generated by supersymmetric localization on the four-sphere reproduces the perturbative calculation of the $\frac14$-BPS Wilson loop (\ref{eq:1/4 Wilson loop}) up to three loop  in $\mathcal{N}=2$ SYM theories with conformal matter in arbitrary representation $\cR$ of SU($N$). Moreover,   in  the $\mathbf{E}$ theory, where conformal matter transform in the rank-two symmetric and antisymmetric representation, the matrix model prediction reveals that the observable coincides with the result of $\mathcal{N}=4$ SYM up to three loops in perturbation theory (\ref{eq:observable in E theory}) and the two observables differ only at order $g^8$ by a term proportional to $\zeta(5)$.  In this section, we show that this prediction is consistent with the perturbative calculation in flat space.

Firstly, using eq.\1(\ref{W2kis def}), we write the generic four-loop correction ($\cW_8$) to the expectation value of the operator (\ref{eq:1/4 Wilson loop}) as follows \begin{equation}
	\cW_{8}= \cW_8^{\mathcal{N}=4} + \cW_{8}^\prime \ ,
\end{equation} where the first term on the right-hand side can be obtained by expanding eq.\1(\ref{eq:vew1/4 N4 and its expansion}) up to order $g^8$, and we recall that the primed contribution denotes the  difference between the diagrams with internal lines  in the representation $\cR$ and the analogous one in the adjoint representation, i.e. $\cW_{8}^\prime=\cW^\cR_8-\cW^{\rm Adj}_8$. 
For an arbitrary representation $\cR$, $\cW_{8}^\prime$ involves several contributions. However, in the $\mathbf{E}$ theory the analysis of the Feynman diagrams drastically simplifies since the two-loop corrections to the adjoint scalar and gauge field propagator (\ref{eq:2-loop diagrammatic corrections difference loop scalar}) vanishes because they are proportional to the coefficient $\cC_{4}^\prime$ (\ref{eq:vev at three-loop def}), i.e.  \begin{equation}
	\label{eq:two-loop in the E theory}
		\mathord{
		\begin{tikzpicture}[scale=0.55, baseline=-0.65ex]
			\filldraw[color=gray!80, fill=gray!15](0,0) circle (1);	
			\draw [black] (0,0) circle [radius=1cm];
			\draw [black, thick, dashed] (0,0) circle [radius=0.9cm];
			\begin{feynman}
				\vertex (A) at (-2,0);
				\vertex (C) at (-1,0);
				\vertex (C1) at (0.1,0) {\text{\footnotesize 2-loop }\tiny} ;
				\vertex (B) at (1, 0);
				\vertex (D) at (2, 0);
				\diagram*{
					(A) -- [photon] (C),
					(B) --[photon] (D),
					(A) -- [fermion] (C),
					(B) --[fermion] (D),
				};
			\end{feynman}
		\end{tikzpicture} 
	} \propto \cC_{4}^\prime =0  \ .
\end{equation}  This means that all the interaction corrections contributing to $\cW^\prime_8$ characterized by subdiagrams involving  the propagators (\ref{eq:two-loop in the E theory}) trivially vanish in these theories. Moreover, in this theory also the diagrams involving  the \textit{irreducible} two-loop corrections to the gauge-scalar pure-gauge vertices in the difference method vanish. In fact, these corrections can be organized in terms of two different classes of diagrams which take the following (schematic) form \cite{Billo:2019fbi}
\begin{equation}
		\mathord{
		\begin{tikzpicture}[radius=2.cm, baseline=-0.65ex, scale=0.6]
			\draw [black] (0,0) circle [radius=0.8cm];
			\draw [black, dashed] (0,0) circle [radius=0.7cm];
			\begin{feynman}
				\vertex (A) at (0,2);
				\vertex (C) at (0,0.8);
				\vertex (D) at (-1.5, -1.3);
				\vertex (B) at (-0.7, -0.4);
				\vertex (B1) at (0.7,-0.4);
				\vertex (B2) at (1.5,-1.3);
				\vertex (C3) at (0,1.6);
				\vertex (C4) at (0.7, -0.55);
				\diagram*{
					(A) -- [photon] (C),
					(B) --[ anti fermion] (D),
					(B) --[photon] (D),
					(B1) --[ fermion] (B2),
					(B1) --[photon] (B2),
					(C3) -- [photon, half left] (C4), 
					(C3) -- [fermion, half left] (C4)
				};
			\end{feynman}
		\end{tikzpicture} 
	} \propto \mathrm{i} f^{abc} \beta_0 \ , \quad \quad 	\mathord{
	\begin{tikzpicture}[radius=2.cm, baseline=-0.65ex, scale=0.6]
		\draw [black] (0,0) circle [radius=0.8cm];
		\draw [black, dashed] (0,0) circle [radius=0.7cm];
		\begin{feynman}
			\vertex (A) at (0,2);
			\vertex (C) at (0,0.8);
			\vertex (D) at (-1.5, -1.3);
			\vertex (B) at (-0.7, -0.4);
			\vertex (B1) at (0.7,-0.4);
			\vertex (B2) at (1.5,-1.3);
			\vertex (C3) at (-0.8,0);
			\vertex (C4) at (0.8, 0);
			\diagram*{
				(A) -- [photon] (C),
				(B) --[ anti fermion] (D),
				(B) --[photon] (D),
				(B1) --[ fermion] (B2),
				(B1) --[photon] (B2),
				(C3) -- [plain] (C4), 
				(C3) -- [photon] (C4)
			};
		\end{feynman}
	\end{tikzpicture} 
} \propto \mathrm{i}f^{abc} \cC_4^\prime \ .
\end{equation}
Importantly, for an arbitrary (conformal) representation $\cR$, the second diagram does not vanishes, while it can be discarded in the $\mathbf{E}$ theory since it is proportional to the coefficient $\cC_{4}^\prime$. This means that the only class of corrections contributing to $\cW^\prime_8$ consists of the following single exchange diagrams 
\begin{equation}
	\label{eq:four-loop diagrams}
	g^8	\cW^{\prime}_{8} =  	\mathord{ \begin{tikzpicture}[baseline=-0.65ex,scale=0.6]
		\draw
		(-3,-2) coordinate (a) 
		-- (0,2) coordinate (b) 
		-- (3,-2) coordinate (c) pic["$\delta$",  draw, angle eccentricity=1.2, angle radius=1cm]
		{angle=a--b--c}; 
		\begin{feynman}
			\vertex (C) at (1.2,0.4);
			\vertex (D) at (2.9,-1.8);
			\newcommand\tmpda{0.4cm}
			\newcommand\tmpdb{2.cm}
			\diagram*{
				(C) -- [photon, half left] (D),
				(C) -- [plain, half left, with arrow=\tmpdb] (D),
				(C) -- [plain, half left, with arrow=\tmpda] (D),
			};
		\end{feynman}
		\filldraw[color=gray!80, fill=gray!15](3,0) circle (0.9);	
		\draw [black] (3,0) circle [radius=0.9cm];
		\draw [black,dashed,thick] (3,0) circle [radius=0.8cm];
		\begin{feynman}
			\vertex (c)  at (3.1,0) {\text{\footnotesize 3-loop }\normalsize} ;
		\end{feynman}
	\end{tikzpicture}
}  \ + \  	\mathord{ \begin{tikzpicture}[baseline=-0.65ex,scale=0.6]
		\draw
		(-3,-2) coordinate (a) 
		-- (0,2) coordinate (b) 
		-- (3,-2) coordinate (c) pic["$\delta$",  draw, angle eccentricity=1.2, angle radius=1cm]
		{angle=a--b--c}; 
		\begin{feynman}
			\vertex (C) at (-2.6,-1.5);
			\vertex (D) at (2.6,-1.5);
			\newcommand\tmpda{0.4cm}
			\newcommand\tmpdb{2.6cm}
			\diagram*{
				(C) -- [photon] (D),
				(C) -- [plain,with arrow=\tmpda] (D),
				(C) -- [plain,with arrow=\tmpdb] (D),
			};
		\end{feynman}
		\filldraw[color=gray!80, fill=gray!15](0,-1.4) circle (0.9);	
		\draw [black] (0,-1.4) circle [radius=0.9cm];
		\draw [black,dashed,thick] (0,-1.4) circle [radius=0.8cm];
		\begin{feynman}
			\vertex (c)  at (0.1,-1.4) {\text{\footnotesize 3-loop }\normalsize} ;
		\end{feynman}
	\end{tikzpicture}
}  \ .
\end{equation} In the previous expression, the internal bubble in the dashed/continuos line notation represents the three-loop correction to the adjoint scalar and gauge field propagators in the difference method. Since we are working in a superconformal theory and eq.\1(\ref{eq:four-loop diagrams}), where the result is regular in four dimensions,  the only non-trivial part of the three-loop corrections to the propagators that can contribute to eq.\1(\ref{eq:four-loop diagrams}) is the regular one. This was explicitly calculated in \cite{Billo:2019fbi} by the superfield formalism, where it was showed that (see in particular eq.\1(4.26))
\begin{equation}
	\mathord{
	\begin{tikzpicture}[scale=0.55, baseline=-0.65ex]
		\filldraw[color=gray!80, fill=gray!15](0,0) circle (1);	
		\draw [black] (0,0) circle [radius=1cm];
		\draw [black, thick, dashed] (0,0) circle [radius=0.9cm];
		\begin{feynman}
			\vertex (A) at (-2,0);
			\vertex (C) at (-1,0);
			\vertex (C1) at (0.1,0) {\text{\footnotesize 3-loop }\tiny} ;
			\vertex (B) at (1, 0);
			\vertex (D) at (2, 0);
			\diagram*{
				(A) -- [fermion] (C),
				(B) --[fermion] (D),
			};
		\end{feynman}
	\end{tikzpicture} 
} = g^6 \dfrac{f^{(3)}_{\mathbf{E}}}{x_{12}^2} \ , \quad \text{where} \quad f^{(3)}_{\mathbf{E}}(d) = -\dfrac{15\zeta(5)\cC^\prime_6}{2^{10}\pi^8}+\cO(d-4) \ ,
\end{equation} and we recall that the coefficient $\cC_6^\prime$ is defined in eq.\1(\ref{eq:c6}). By supersymmetry, upon insertion in the diagrams (\ref{eq:four-loop diagrams}), the analogous correction to the gauge field propagator coincides with the previous expression up to the metric tensor $\delta_{\mu \nu}$. This means that we can obtain the final expression of eq.\1(\ref{eq:four-loop diagrams}) by the replacement $f(d)\to f^{(3)}_\mathbf{E}(d)$ in eq.\1(\ref{eq:net result w22}), i.e. \begin{equation}
\label{eq:interaction term in the E theory}
g^8\cW^\prime_8 =  g^2 C_F f^{(3)}(d)\pi^2 \alpha^2
\Big|_{d=4}=-g^8 \dfrac{C_F15\zeta(5)\cC_6^\prime}{2^{10}\pi^6}
\alpha^2 \ .
\end{equation} The previous expression matches the matrix model prediction for the $\mathbf{E}$ model (\ref{eq:observable in E theory}).

\bibliographystyle{JHEP}
\bibliography{biblio}
	
\end{document}